\documentclass[aps,prc,amsmath,amssymb,superscriptaddress,groupedaddress]{revtex4}

\usepackage{graphicx}
\usepackage{dcolumn}
\usepackage{bm}
\usepackage{enumerate}
\usepackage{amsmath}
\usepackage{amssymb}
\usepackage{tabularx}
\usepackage{mathtools}
\usepackage{xcolor}
\usepackage{ctable}
\usepackage{rotating}
\usepackage{wrapfig}

\begin{document}

%\preprint{APS/PRC - Nuclear Reaction}

\title{Alleviating the inconsistencies in modelling decay of fissile compound
nuclei}

\author{Tathagata Banerjee}
\email{he.tatha@gmail.com}
\altaffiliation{Presently at Department of Nuclear Physics, Research School of
Physics and Engineering, The Australian National University, Canberra ACT 0200,
Australia.}
\affiliation{Nuclear Physics Group, Inter University Accelerator Centre,
Aruna Asaf Ali Marg, Post Box 10502, New Delhi 110067, India}
\author{S. Nath}
\affiliation{Nuclear Physics Group, Inter University Accelerator Centre,
Aruna Asaf Ali Marg, Post Box 10502, New Delhi 110067, India}
\author{Santanu Pal}
\altaffiliation{Formerly with Physics Group, Variable Energy Cyclotron Centre,
1/AF Bidhan Nagar, Kolkata 700064, India.}
\affiliation{Nuclear Physics Group, Inter University Accelerator Centre,
Aruna Asaf Ali Marg, Post Box 10502, New Delhi 110067, India}

\date{\today}

\begin{abstract}
Despite remarkable success of the statistical model (SM) in describing decay of
excited compound nuclei (CN), reproduction of the observables from heavy
ion-induced fusion-fission reactions is often quite challenging. Ambiguities in
choosing the input parameters, lack of clarity about inclusion of various
physical effects in the model and contradictory requirements of input
parameters while describing different observables from similar reactions are
among the major difficulties of modelling decay of fissile CN. 
This work attempts to overcome the existing inconsistencies by inclusion of
important physical effects in the model and through a systematic analysis of a
large set of data over a wide range of CN mass ($A_{\textrm{CN}}$). 
The model includes shell effect in the level density (LD) parameter, shell
correction in the fission barrier ($B_{\textrm{f}}$), effect of the orientation
degree of freedom of the CN spin ($K_{\textrm{or}}$), collective enhancement of
level density (CELD) and dissipation in fission. Input parameters are not tuned
to reproduce observables from specific reaction(s) and the reduced dissipation
coefficient ($\beta$) is treated as the only adjustable parameter. Calculated
evaporation residue (ER) cross sections ($\sigma_{\textrm{ER}}$), fission cross
sections ($\sigma_{\textrm{fiss}}$) and particle, \textit{i.e.} neutron, proton
and $\alpha$-particle, multiplicities are compared with data covering
$A_{\textrm{CN}} =$ 156 \textendash 248. 
The model produces reasonable fits to ER and fission excitation functions for
all the reactions considered in this work. Pre-scission neutron
multiplicities ($\nu_{\textrm{pre}}$) are underestimated by the calculation
beyond $A_{\textrm{CN}} \sim 200$. An increasingly higher value of $\beta$, in
the range of 2 \textendash 4 $\times 10^{21}$ s$^{-1}$, is required to
reproduce the data with increasing $A_{\textrm{CN}}$. Proton and
$\alpha$-particle multiplicities, measured in coincidence with both ERs and
fission fragments, are in qualitative agreement with model predictions.
The present work mitigates the existing inconsistencies in modelling
statistical decay of the fissile CN to a large extent. Contradictory
requirements of fission enhancement, by scaling down the fission barrier, to
reproduce $\sigma_{\textrm{ER}}$ or $\sigma_{\textrm{fiss}}$ and fission
suppression, by introducing dissipation in the fission channel, to reproduce
$\nu_{\textrm{pre}}$ for similar reactions have now become redundant. There are
scopes for further refinement of the model, as is evident from the mismatch
between measured and calculated particle multiplicities in a few cases.
\end{abstract}

%\pacs{27.80.+w,25.70.Jj,24.60.Dr}

\maketitle

\section{Introduction}
\label{intro}
Concepts of statistical mechanics have been applied to describe the decay of
an excited compound nucleus (CN) since the same was hypothesized by Bohr
\cite{Bohr1936} as a mono-nucleus fully equilibrated in all degrees of freedom
with no memory of its formation except the conserved quantities such as energy
and angular momentum. The resulting formalism, termed generally as the
statistical model (SM) of decay of the CN has been quite successfully employed
in reproducing observables from fusion reactions over the last several decades.
There are, yet, certain ambiguities and inconsistencies in interpreting results
from heavy ion-induced fusion-fission reactions
\cite{TathaPLB2018,Schmitt2014}. We start with a few of the open questions:

\begin{enumerate}

\item Simultaneous reproduction of evaporation residue (ER) cross section
($\sigma_{\textrm{ER}}$), fission cross section ($\sigma_{\textrm{fiss}}$),
pre-scission neutron, proton and alpha-particle multiplicities
($\nu_{\textrm{pre}}$, $\pi_{\textrm{pre}}$ and $\alpha_{\textrm{pre}}$) is not
successful till date. As a consequence, it has become an accepted practice to 
reproduce the measured quantities by tuning the SM parameters \textit{viz.} 
level density parameters at ground state and saddle, a scaling factor for the 
fission barrier, a pre-saddle delay and saddle-to-scission transition time 
\cite{Delag1977,Videbaek1977,Ward1983,LestonePRL1993,Fineman1994,Charity2010,Nitto2011}. 
Several combinations of those SM parameters could reproduce the data 
\cite{Mancusi2010,McCalla2008,Lestone2009}. One naturally wonders if there is a
way to reproduce data without \textit{ad hoc} manipulation of the SM
parameters. 

\item While a speeding up of fission, by means of reducing the fission barrier,
was required to reproduce measured $\sigma_{\textrm{ER}}$
\cite{Brinkmann1994,Sagaidak2009,Varinderjit2014}, a slowing down of fission
was necessary to reproduce experimental $\nu_{\textrm{pre}}$
\cite{Hinde1986,Newton1988}. Does this contradiction point to an inadequate
modelling of CN decay?

\item The SM fails to reproduce the particle multiplicities measured in
coincidence with fission fragments (FFs) ($\nu_{\textrm{pre}}$,
$\pi_{\textrm{pre}}$ and $\alpha_{\textrm{pre}}$) and with ERs
($\nu_{\textrm{ER}}$, $\pi_{\textrm{ER}}$ and $\alpha_{\textrm{ER}}$),
simultaneously \cite{Vardaci2010}. Can this inconsistency be overcome?

\item Though disentangling pre-scission particle emissions from post-scission
emissions is possible experimentally, the same is fraught with difficulties in
case of the particles emitted in pre-saddle and post-saddle regimes
\cite{Hilscher1992}. One can estimate the pre-scission dissipation coefficient
from the analysis of light particle spectra and giant dipole resonance (GDR)
$\gamma$-ray multiplicities
\cite{Hilscher1992,Paul1994,Dioszegi2000,Frobrich1998,Shaw2000}. But, to
acquire a precise and reliable information about the pre-saddle dissipation
coefficient, one must employ those experimental signatures which are uniquely
sensitive to the pre-saddle regime only, such as, $\sigma_{\textrm{ER}}$
\cite{Laveen2015} and ER spin distribution \cite{Mohanto2012}. Different
combinations of pre-saddle and saddle-to-scission dissipation coefficients
succeeded in interpreting particle multiplicity data
\cite{Paul1994,Dioszegi2000,Shaw2000,Ramachandran2006}. Is it possible to
determine the pre-saddle dissipation coefficient accurately?

\end{enumerate}

In this article, we present a consistent SM description of observables from
heavy ion-induced fusion-fission reactions with the reduced dissipation
coefficient ($\beta$) as the only adjustable parameter. We shall apply shell
effect to the level density (LD) and shell correction to the fission barrier
($B_{\textrm{f}}$). Effects of orientation degree of freedom of CN spin
($K_{\textrm{or}}$) and collective enhancement of LD (CELD) are also included
in the present model. We aim in this work to include all the effects which
impact fission and various evaporation widths in order to fit a broad range of
data with a minimum of adjustment of input parameters. A shorter version of the
results obtained from SM calculations with the aforementioned effects has been
reported earlier \cite{TathaPLB2018}. Here we present results for a larger set
of systems and for a wider range of experimental observables.

\begin{table*}
\begin{center}
\caption{\label{tab:table1} List of reactions for which measured
$\sigma_{\textrm{ER}}$, $\sigma_{\textrm{fiss}}$ and $\nu_{\textrm{pre}}$ are
compared with model predictions. Comparisions are shown in Fig. \ref{FIGURE3}
\textendash Fig. \ref{FIGURE6}.}
\begin{tabular}{l l l l l l l}
\hline
\hline
Reaction & CN & Ref. & &Reaction & CN & Ref.\\
\hline
$^{12}$C+$^{144}$Sm & $^{156}$Er & \cite{Janssens1986} & & $^{64}$Ni+$^{92}$Zr & $^{156}$Er & \cite{Janssens1986,Wolfs1989}\\
$^{12}$C+$^{158}$Gd & $^{170}$Yb & \cite{Zebelman1973,GavronPRL1981} & & $^{16}$O+$^{154}$Sm & $^{170}$Yb & \cite{Leigh1995,Zebelman1973,Hinde1986,HindeRapid1988,Hinde1992}\\
$^{20}$Ne+$^{150}$Nd & $^{170}$Yb & \cite{Sarantites1976,Zebelman1974,Halbert1978,GavronPRL1981} & & $^{4}$He+$^{188}$Os & $^{192}$Pt & \cite{Navin2004,IgnatyukJNucl1975,Schmitt2003}\\
$^{16}$O+$^{176}$Yb & $^{192}$Pt & \cite{Tapan2016,Schmitt2003} & & $^{16}$O+$^{181}$Ta & $^{197}$Tl & \cite{Devendra2009,Videbaek1977,Ogihara1990,Bivash2001,Hardev2007}\\
$^{19}$F+$^{178}$Hf & $^{197}$Tl & \cite{Hardev2007} & & $^{16}$O+$^{184}$W & $^{200}$Pb & \cite{Leigh1988,Shidling2006,Bemis1986,Forster1987,Hinde1992}\\
$^{19}$F+$^{181}$Ta & $^{200}$Pb & \cite{Hinde1986,Charity1986,Newton1988} & & $^{30}$Si+$^{170}$Er & $^{200}$Pb & \cite{Hinde1982,Gayatri2013,Newton1988}\\
$^{1}$H+$^{209}$Bi & $^{210}$Po & \cite{Beyec1970,Gadioli1969,Khodai1966,Ignatyuk1984,Shigaev1973,Cheifetz1970} & & $^{4}$He+$^{206}$Pb & $^{210}$Po & \cite{Bimbot1969,Khodai1966,Schmitt2003}\\
$^{12}$C+$^{198}$Pt & $^{210}$Po & \cite{Shrivastava1999,Plicht1983,Golda2013} & & $^{18}$O+$^{192}$Os & $^{210}$Po & \cite{Charity1986,Plicht1983,Newton1988}\\
$^{12}$C+$^{204}$Pb & $^{216}$Ra & \cite{Hinde2002,Sagaidak2003,Hardev2008} & & $^{19}$F+$^{197}$Au & $^{216}$Ra & \cite{Hinde2002,Hardev2008}\\
$^{30}$Si+$^{186}$W & $^{216}$Ra & \cite{Hinde2002} & & $^{1}$H+$^{238}$U & $^{239}$Np & \cite{Bate1964,Kandil1976,Cheifetz1970,Strecker1990,Rubchenya2001}\\
$^{7}$Li+$^{232}$Th & $^{239}$Np & \cite{Freies1975,Hinde1989} & & $^{11}$B+$^{237}$Np & $^{248}$Cf & \cite{Liu1996,Kailas1999,Saxena1994}\\
$^{16}$O+$^{232}$Th & $^{248}$Cf &\cite{Back1985,Vandenbosch1986,Nadkarni1999,Saxena1994} & & & & \\
\hline
\hline
\end{tabular}
\end{center}
\end{table*}

In order to compare the predictions of the present SM with data, we choose
those reactions for which non-CN fission (NCNF) is predicted to be
small on the basis of a systematic analysis of ER cross-sections
\cite{Tatha2015}. We consider here population of $^{156}$Er, $^{170}$Yb,
$^{192}$Pt, $^{197}$Tl, $^{200}$Pb, $^{210}$Po, $^{216}$Ra, $^{239}$Np and
$^{248}$Cf CN, each by at least two different entrance channels. The list of
reactions is presented in Table \ref{tab:table1}. We compare
$\sigma_{\textrm{ER}}$, $\sigma_{\textrm{fiss}}$ and $\nu_{\textrm{pre}}$
($\nu_{\textrm{ER}}$, in a few cases) of these reactions with the predictions
of the present model. We further compare $\pi_{\textrm{ER}}$,
$\pi_{\textrm{pre}}$, $\alpha_{\textrm{ER}}$ and $\alpha_{\textrm{pre}}$
of eight reactions having $A_{\textrm{CN}} =$ 158 \textendash 225 with the
model predictions. Table \ref{tab:table2} contains the list of these reactions.

\begin{table}
\begin{center}
\caption{\label{tab:table2}List of reactions for which measured $\sigma_{\textrm{ER}}$, proton and
$a$-particle multiplicities are compared with model predictions. Comparisions
are shown in Fig. \ref{FIGURE7} and Fig. \ref{FIGURE8}.}
\begin{tabular}{l l l l l l l}
\hline
\hline
Reaction & CN & Ref. & &Reaction & CN & Ref.\\
\hline
$^{32}$S+$^{126}$Te & $^{158}$Er & \cite{Nadotchy2010} & & $^{19}$F+$^{159}$Tb & $^{178}$W & \cite{Ikezoe1994}\\
$^{28}$Si+$^{165}$Ho & $^{193}$Tl & \cite{Fineman1994} & & $^{19}$F+$^{181}$Ta & $^{200}$Pb & 
\cite{Hinde1986,Caraley2000,Ikezoe1992,Fabris1994}\\
$^{16}$O+$^{197}$Au & $^{213}$Fr & \cite{Ikezoe1992,Brinkmann1994,Fineman1994} & & $^{19}$F+$^{197}$Au & $^{216}$Ra 
& \cite{Ikezoe1992}\\
$^{16}$O+$^{208}$Pb & $^{224}$Th & \cite{Fineman1994,Brinkmann1994,Morton1995} & & $^{28}$Si+$^{197}$Au & $^{225}$Np 
& \cite{Ikezoe1992}\\
\hline
\hline
\end{tabular}
\end{center}
\end{table}

The article is organized as follows. The various ingredients of the SM
calculations and the results are presented in section \ref{ModRes} followed by
a discussion in Sec. \ref{Dis}. We summarize and conclude in Sec. \ref{Conc}.

\section{Statistical Model Calculations}
\label{ModRes}
\subsection{The Model}

\begin{figure}[ht!]
\includegraphics[width=0.48\textwidth,trim=0.0cm 0.0cm 0.0cm 0.0cm,clip]{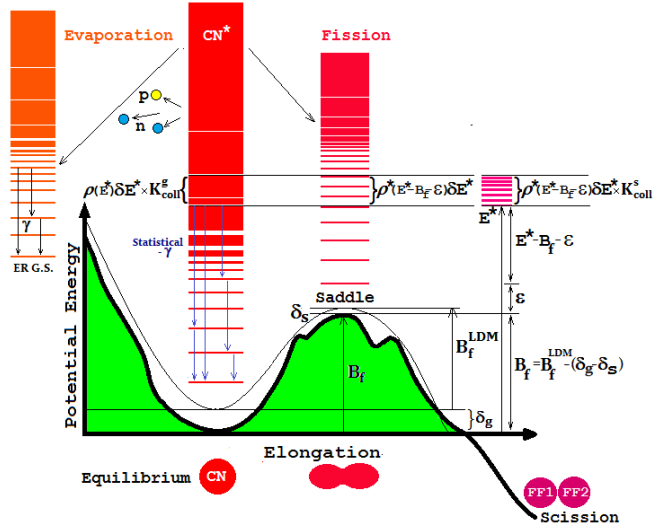}
\caption{\label{FIGURE1} Potential energy as a function of elongation. The
fission barrier ($B_{\textrm{f}}$) and the density of states ($\rho(E^{*})$)
change when shell corrections and CELD are taken into account, leading to
modification of the Bohr-Wheeler fission width. $\rho(E^{*})\delta E^{*}$ is
the number of quantum states between energies $E^{*}$ and
$E^{*} + \delta E^{*}$ at the ground state (\textit{i.e.} local minima with
zero potential energy). The number of quantum states at the saddle point, with
inclusion of shell corrections, would be
$\rho(E^{*} - B_{\textrm{f}} - \epsilon) \delta E^{*}$, where $\epsilon$ is the
associated kinetic energy. With the incorporation of CELD, the same would be
modified to
$\rho(E^{*} - B_{\textrm{f}} - \epsilon)\delta E^{*} \times K_{\textrm{coll}}^{\textrm{s}}$
and also the number of states available at the ground state would become
$\rho(E^{*}) \delta E^{*} \times K_{\textrm{coll}}^{\textrm{g}}$.
$\delta_{\textrm{g}}$ and $\delta_{\textrm{s}}$ are shell correction energies
at the ground state and the saddle point, respectively. See text for details.}
\end{figure}

\begin{figure}[ht!]
\includegraphics[width=0.45\textwidth,trim=0.0cm 0.0cm 0.0cm 0.0cm,clip]{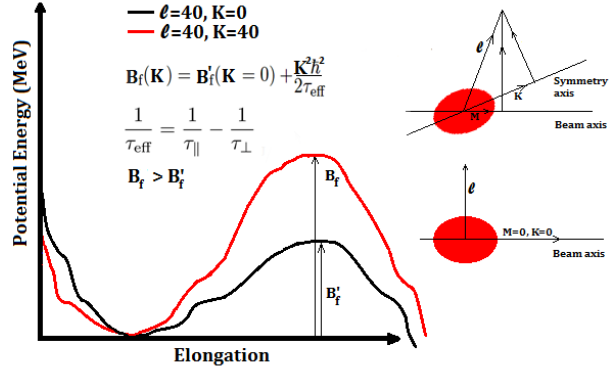} 
\caption{\label{FIGURE2} The effect of $K$ degree of freedom
($K_{\textrm{or}}$) on $B_{\textrm{f}}$. The increased value of
$B_{\textrm{f}}$ would reduce the fission width. See text for details.} 
\end{figure}

\begin{figure*}[ht!]
\begin{center}
\includegraphics[width=\textwidth,trim=0.0cm 0.0cm 0.0cm 0.0cm,clip]{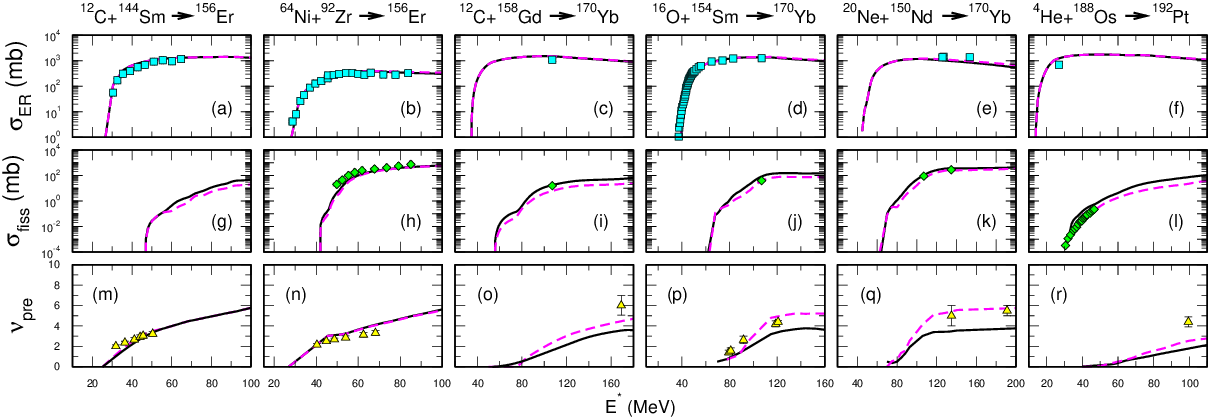}
\caption{\label{FIGURE3} Comparison of measured $\sigma_{\textrm{ER}}$,
$\sigma_{\textrm{fiss}}$ and $\nu_{\textrm{pre}}$ with SM predictions for the
reaction $^{12}$C+$^{144}$Sm \cite{Janssens1986}, $^{64}$Ni+$^{92}$Zr
\cite{Janssens1986,Wolfs1989}, $^{12}$C+$^{158}$Gd
\cite{Zebelman1973,GavronPRL1981}, $^{16}$O+$^{154}$Sm
\cite{Leigh1995,Zebelman1973,Hinde1986,HindeRapid1988,Hinde1992},
$^{20}$Ne+$^{150}$Nd
\cite{Sarantites1976,Zebelman1974,Halbert1978,GavronPRL1981} and
$^{4}$He+$^{188}$Os \cite{Navin2004,IgnatyukJNucl1975,Schmitt2003}. Neutron
multiplicities of $^{12}$C+$^{144}$Sm and $^{64}$Ni+$^{92}$Zr were measured and
calculated in coincidence with ER. Continuous (black) lines indicate the SM
predictions including shell correction in both $B_{\textrm{f}}$ and LD, CELD
and $K$-orientation effects. Dashed (magenta) lines show results with all the
aforementioned effects and $\beta = 2 \times 10^{21}$ s$^{-1}$.}
\end{center}
\end{figure*}
\begin{figure*}[ht!]
\begin{center}
\includegraphics[width=\textwidth,trim=0.0cm 0.0cm 0.0cm 0.0cm,clip]{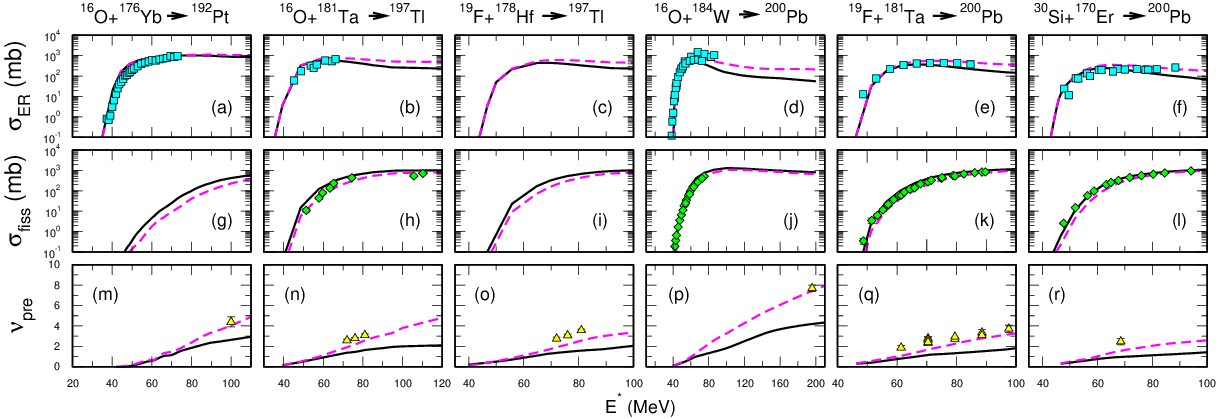}
\caption{\label{FIGURE4} Comparison of measured $\sigma_{\textrm{ER}}$,
$\sigma_{\textrm{fiss}}$ and $\nu_{\textrm{pre}}$ with SM predictions for the
reactions $^{16}$O+$^{176}$Yb \cite{Tapan2016,Schmitt2003}, $^{16}$O+$^{181}$Ta
\cite{Devendra2009,Videbaek1977,Ogihara1990,Bivash2001,Hardev2007},
$^{19}$F+$^{178}$Hf \cite{Hardev2007}, $^{16}$O+$^{184}$W
\cite{Leigh1988,Shidling2006,Bemis1986,Forster1987,Hinde1992},
$^{19}$F+$^{181}$Ta \cite{Hinde1986,Charity1986,Newton1988} and
$^{30}$Si+$^{170}$Er \cite{Hinde1982,Gayatri2013,Newton1988}. Continuous
(black) lines indicate the SM predictions including shell correction in both
$B_{\textrm{f}}$ and LD, CELD and $K$-orientation effects. Dashed (magenta)
lines show results with all the aforementioned effects and
$\beta = 2 \times 10^{21}$ s$^{-1}$.}
\end{center}
\end{figure*}

In the SM calculations, the time evolution of a CN (an event) is followed over
small time steps and the fate of the CN at each time step is decided by a Monte
Carlo sampling of the decay widths of various channels. The CN can follow
various decay routes depending upon the relative probabilities of different
decay channels. We consider fission along with emission of neutrons,
protons, $\alpha$-particles and $\gamma$-rays as the decay channels of a CN. A
CN can either undergo fission with or without preceding evaporation of
particles and photons or reduce to an ER. The final results for different
observables are obtained as averages over a large ensemble of events. We assume
the dominant fission mode to be symmetric and the fission width is obtained
from the transition-state model of fission due to Bohr and Wheeler
\cite{Bohr1939}. The particle and $\gamma$ decay widths are obtained from the
Weisskopf formula as given in Ref. \cite{Frobrich1998}.

The evaporation and fission widths depend upon the spin of the CN and hence the
spin distribution of CN, formed in a fusion-fission reaction, is required in SM
calculations. We obtain this distribution by assuming that the whole of the
incident flux in the entrance channel is absorbed to form a CN. Therefore we
consider the fusion cross-section the same as the capture cross-section and
obtain the spin distribution of the CN from the coupled-channels code 
\textsc{ccfull} \cite{Hagino1999} using coupling constants and excitation
energies of the low-lying collective states of both the projectile and the
target nucleus. The CN spin distribution, thus obtained, is used as input to
the SM calculation.

The fission barrier in the present calculation is obtained by including shell
correction in the liquid-drop nuclear mass. The macroscopic part of the fission
barrier is given by the finite-range liquid drop model (FRLDM) which was
obtained earlier by fitting the systematic behaviour of ground state masses and
fission barriers at low angular momentum for nuclei over a wide range of masses
\cite{Sierk1986}. The shell correction term $\delta$ is defined as the
difference between the experimental and the liquid-drop model (LDM) masses,
\begin{equation}
\label{ShellCorr}
\delta = M_{\textrm{exp}} - M_{\textrm{LDM}} .
\end{equation}
\noindent           
The full fission barrier $B_{\textrm{f}}(\ell)$ of a nucleus carrying angular
momentum $\ell$ is then given as
\begin{equation}
\label{barrier}
B_{\textrm{f}}(\ell) = B_{\textrm{f}}^{\textrm{LDM}}(\ell) -
\left( \delta_{\textrm{g}} - \delta_{\textrm{s}} \right)
\end{equation}
\noindent           
where, $B_{\textrm{f}}^{\textrm{LDM}}(\ell)$ is the angular momentum dependent
LDM fission barrier \cite{Sierk1986} and $\delta_{\textrm{g}}$ and
$\delta_{\textrm{s}}$ are the shell correction energies for the ground-state
and saddle configurations, respectively. The shell corrections at ground state
and saddle are obtained following the recipe given in Ref. \cite{Myers1966} for
including deformation dependence in shell correction energy. This yields a
negligible shell correction at large deformations while the full shell
correction is applied at zero deformation. A schematic representation of the
shell effect on available phase space at ground state and saddle is given in
Fig. \ref{FIGURE1}. 

The shell structure in nuclear single-particle levels also influences the
nuclear level density which is used to calculate various decay widths of the
compound nucleus. Ignatyuk et al. \cite{Ignatyuk1975} proposed a level density
parameter $a$ which includes an intrinsic excitation energy ($E^{*}$) dependent
shell effect term and is given as 
\begin{equation}
a(E^{*}) = \tilde{a} \left[1 + \frac{1 - \exp \left (-\frac{E^{*}}{E_{\textrm{D}}} \right )}{E^{*}} \delta \right] .
\end{equation} 
\noindent
Here, $E_{\textrm{D}}$ is a parameter which determines the rate at which the
shell effect decreases with increase of $E^{*}$. The above form of the level
density parameter used in the present work exhibits shell effects at low
excitation energies and goes over to its asymptotic value at high excitation
energies. The following asymptotic shape-dependent level density parameter is
taken from the work of Reisdorf \cite{Reisdorf1981},
\begin{equation}
\tilde{a} = 0.04543 r_{0}^{3}A + 0.1355 r_{0}^{2}A^{\frac{2}{3}}B_{\textrm{s}}+0.1426r_{0}A^{\frac{1}{3}}B_{\textrm{k}}
\end{equation}
\noindent
where $A$ is the nuclear mass number, $r_{0}$ is the nuclear radius parameter
and $B_{\textrm{s}}$ and $B_{\textrm{k}}$ are respectively the surface and
curvature terms of the liquid drop model. The values of $r_{0}$ and
$E_{\textrm{D}}$ are fixed by fitting the available $s$-wave neutron resonance
spacings \cite{Reisdorf1981}.

\begin{figure*}[ht!]
\begin{center}
\includegraphics[width=\textwidth,trim=0.0cm 0.0cm 0.0cm 0.0cm,clip]{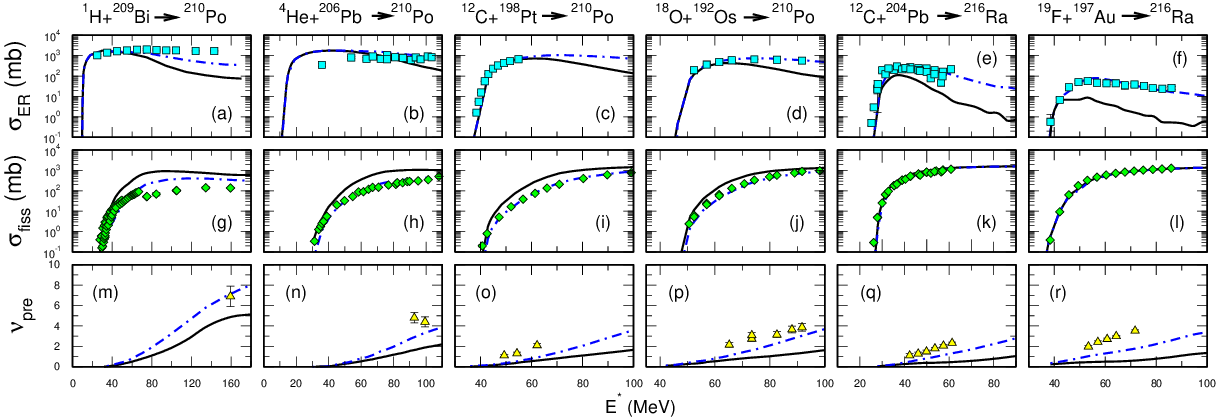}
\caption{\label{FIGURE5} Comparison of measured $\sigma_{\textrm{ER}}$,
$\sigma_{\textrm{fiss}}$ and $\nu_{\textrm{pre}}$ with SM predictions for the
reactions
$^{1}$H+$^{209}$Bi
\cite{Beyec1970,Gadioli1969,Khodai1966,Ignatyuk1984,Shigaev1973,Cheifetz1970}, 
$^{4}$He+$^{206}$Pb \cite{Bimbot1969,Khodai1966,Schmitt2003}, 
$^{12}$C+$^{198}$Pt \cite{Shrivastava1999,Plicht1983,Golda2013},
$^{18}$O+$^{192}$Os \cite{Charity1986,Plicht1983,Newton1988},
$^{12}$C+$^{204}$Pb ~\cite{Hinde2002,Sagaidak2003,Hardev2008} and 
$^{19}$F+$^{197}$Au ~\cite{Hinde2002,Hardev2008}. 
Continuous (black) lines indicate the SM predictions including shell correction
in both $B_{\textrm{f}}$ and LD, CELD and $K$-orientation effects.
Dash-dotted (blue) lines show results with all the aforementioned effects and
$\beta = 3 \times 10^{21}$ s$^{-1}$.}
\end{center}
\end{figure*}
\begin{figure*}[ht!]
\begin{center}
\includegraphics[width=\textwidth,trim=0.0cm 0.0cm 0.0cm 0.0cm,clip]{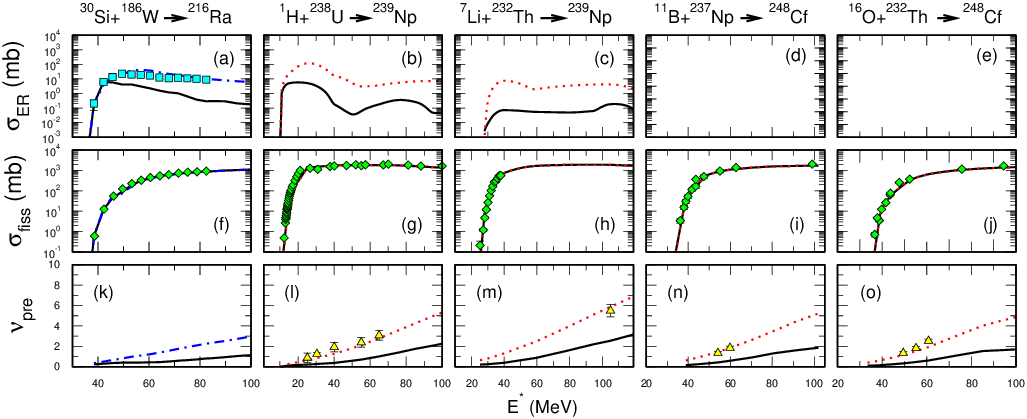}
\caption{\label{FIGURE6} Comparison of measured $\sigma_{\textrm{ER}}$,
$\sigma_{\textrm{fiss}}$ and $\nu_{\textrm{pre}}$ with SM predictions for the
reactions $^{30}$Si+$^{186}$W \cite{Hinde2002}, $^{1}$H+$^{238}$U  
\cite{Bate1964,Kandil1976,Cheifetz1970,Strecker1990,Rubchenya2001},
$^{7}$Li+$^{232}$Th \cite{Freies1975,Hinde1989}, 
$^{11}$B+$^{237}$Np \cite{Liu1996,Kailas1999,Saxena1994} and
$^{16}$O+$^{232}$Th \cite{Back1985,Vandenbosch1986,Nadkarni1999,Saxena1994}.
Continuous (black) lines indicate the SM predictions including shell correction
in both $B_{\textrm{f}}$ and LD, CELD and $K$-orientation effects.
Dash-dotted (blue) and dotted (red) lines show results with all the
aforementioned effects and $\beta = 3 \times 10^{21}$ s$^{-1}$ and
$\beta = 4 \times 10^{21}$ s$^{-1}$, respectvely.}
\end{center}
\end{figure*}

The nuclear level density considered so far corresponds to that of a Fermi gas
with effect of shell structure included at lower excitations. However, residual
interaction in the nuclear Hamiltonian can give rise to correlation among
particle-hole states resulting in collective excitations. The energy levels of
these collective states are often considerably lower than the non-interacting
particle-hole states from which they are formed. Inclusion of collective states
therefore enhances the level density obtained with independent particle model
at low excitation energies. The collective enhancement of level density (CELD)
was considered earlier by Bjornholm, Bohr and Mottelson \cite{BBM1974} where
the collective levels were generated by adding additional degrees of freedom to
those of the Fermi gas. They further argued that the effect of double counting
of states can be neglected since the excitation energy of the particle-hole
states which are involved in the collective states are so high that there are
many more states at these energies in the Fermi gas which do not contribute to
the collective states. The total level density $\rho (E^{*})$ can therefore be
written as \cite{BBM1974}
\begin{equation}
\label{LevDen}
\rho (E^{*}) = K_{\textrm{coll}}(E^{*})\rho_{\textrm{intr}} (E^{*})
\end{equation}
\noindent
where, $\rho_{\textrm{intr}}(E^{*})$ is the intrinsic level density and
$K_{\textrm{coll}}$ is the collective enhancement factor. The rotational and
vibrational enhancement factors are obtained as
\begin{subequations}
\begin{align}
K_{\textrm{rot}} &= \frac{\tau_{\perp}T}{\hbar^{2}} , \\
K_{\textrm{vib}} &= \textit{e}^{0.055 \times A^{\frac{2}{3}}\times T^{\frac{4}{3}}} 
\end{align}
\end{subequations}
\noindent
where $T$ is the nuclear temperature and $\tau_{\perp}$ is the rigid body
moment of inertia perpendicular to the symmetry axis \cite{Ignatyuk1985}. A
smooth transition from $K_{\textrm{vib}}$ to $K_{\textrm{rot}}$ with increasing
quadrupole deformation $|\beta_{2}|$, was obtained by Zagrebaev \textit{et al.}
\cite{Zagrebaev2001}
\begin{subequations}
\begin{equation}
\label{kcol}
K_{\textrm{coll}}(|\beta_{2}|) = [K_{\textrm{rot}}\varphi(|\beta_{2}|)
+K_{\textrm{vib}}(1-\varphi(|\beta_{2}|))]\ f(E^{*})
\end{equation}
\noindent
using a function $\varphi$($|\beta_{2}|$) given as
\begin{equation}
\varphi(|\beta_{2}|) = \left[1+\exp\left(\frac{\beta_{2}^{0}-|\beta_{2}|}
{\Delta\beta_{2}}\right)\right]^{-1}
\end{equation}
\noindent
and the damping of collective effects with increasing excitation is accounted
for by the functional form \cite{Junghans1998}
\begin{equation}
f(E^{\textrm{*}})= \left[1+\exp\left(\frac{E^{*}-E_{\textrm{cr}}}{\Delta E}\right)\right]^{-1}.
\end{equation}
\end{subequations}
\noindent
The values of $\beta_{2}^{0}= 0.15$ and $\Delta\beta_{2}= 0.04$ are taken from
Ref. \cite{Ohta2001}. The values of $E_{\textrm{cr}}$ and $\Delta E$ are taken
as 40 MeV and 10 MeV, respectively, which were obtained by fitting yields from
projectile fragmentation experiments\cite{Junghans1998}. The effect of damping
of CELD with increasing excitation energy has also been experimentally observed
in evaporation spectra of neutrons and high energy photons
\cite{KBanerjee2017,DPandit2018}.

From the transition-state theory of Bohr and Wheeler \cite{Bohr1939}, the
fission width for a nucleus with total excitation energy $E^{*}$ and angular
momentum $\ell$ is given as,
\begin{widetext}
\begin{equation}
\Gamma_{\textrm{f}}^{\textrm{BW}}(E^{*},\ell,K=0) = \frac{1}{2 \pi
\rho_{\textrm{g}}(E^{*})}
\int_{0}^{E^{*}-B_{\textrm{f}}(\ell)}\rho_{\textrm{s}}\left(E^{*}-B_{\textrm{f}}(\ell)-\epsilon\right)d\epsilon
\end{equation}
\end{widetext}
\noindent
where
\begin{equation}
E^{*} = E - E_{\textrm{rot}}(\ell)-E_{\textrm{pair}}
\end{equation}
\noindent
is the intrinsic or thermal part of $E$ and $E_{\textrm{rot}}(\ell)$ and
$E_{\textrm{pair}}$ are the rotational and pairing energies, respectively.
$\rho_{\textrm{g}}$ and $\rho_{\textrm{s}}$ denote the level densities at the
ground state and the saddle configurations, respectively, as given by Eq.
\ref{LevDen}. $B_{\textrm{f}}(\ell)$ is the angular momentum dependent fission
barrier, defined by (Eq. \ref{barrier}).

The above fission width is obtained under the assumption that the orientation
of the angular momentum remains perpendicular to both the reaction plane and
the symmetry axis throughout the course of the reaction. Therefore the angular
momentum component along the symmetry axis ($K$) is set equal to zero in the
above equation. However, the intrinsic nuclear motion can perturb the CN angular
momentum and cause to change its orientation from the initial direction
perpendicular to the symmetry axis \cite{Lestone2009} as schematically
illustrated in Fig. \ref{FIGURE2}. Therefore, $K$ can assume non-zero values
during the fission process. Since the moment of inertia parallel to the symmetry
axis is smaller than that in the perpendicular direction, the rotational energy
at the saddle, and consequently the fission barrier is higher for $K\neq0$
states than that for $K=0$. If a fast equilibration of the $K$-degree of freedom
is assumed, the fission width can be obtained as
\cite{Lestone1999},
\begin{widetext}
\begin{equation}
\label{KOR}
\Gamma_{\textrm{f}}^{\textrm{BW}} (E^{*}, \ell, K) =
\Gamma_{\textrm{f}}^{\textrm{BW}} (E^{*}, \ell, K=0)
\frac{\left(K_{0}\sqrt{2\pi}\right)}{2\ell+1}\ \textrm{erf} 
\left(\frac{\ell+1/2}{K_{0}\sqrt{2}}\right)
\end{equation}
\end{widetext}
\noindent
with
\begin{subequations}
\begin{align}
K_{0}^{2} &= \frac{\tau_{\textrm{eff}}}{\hbar^{2}} T_{\textrm{sad}}, \\
\frac{1}{\tau_{\textrm{eff}}} &= \frac{1}{\tau_{\parallel}} -
\frac{1}{\tau_{\perp}}
\end{align}
\end{subequations}
\noindent
where $T_{\textrm{sad}}$ is the temperature at saddle and $\tau_{\textrm{eff}}$
is the effective moment of inertia. $\tau_{\perp}$ and $\tau_{\parallel}$ are
the moments of inertia at saddle of the nucleus perpendicular to and about the
nuclear symmetry axis. The above definition of $K_{0}^{2}$ from the transition
state model of fission explains the fission fragment angular distribution
satisfactorily. erf(x) denotes the error function. 

\begin{figure*}[ht!]
\begin{center}
\includegraphics[width=\textwidth,trim=0.0cm 0.0cm 0.0cm 0.0cm,clip]{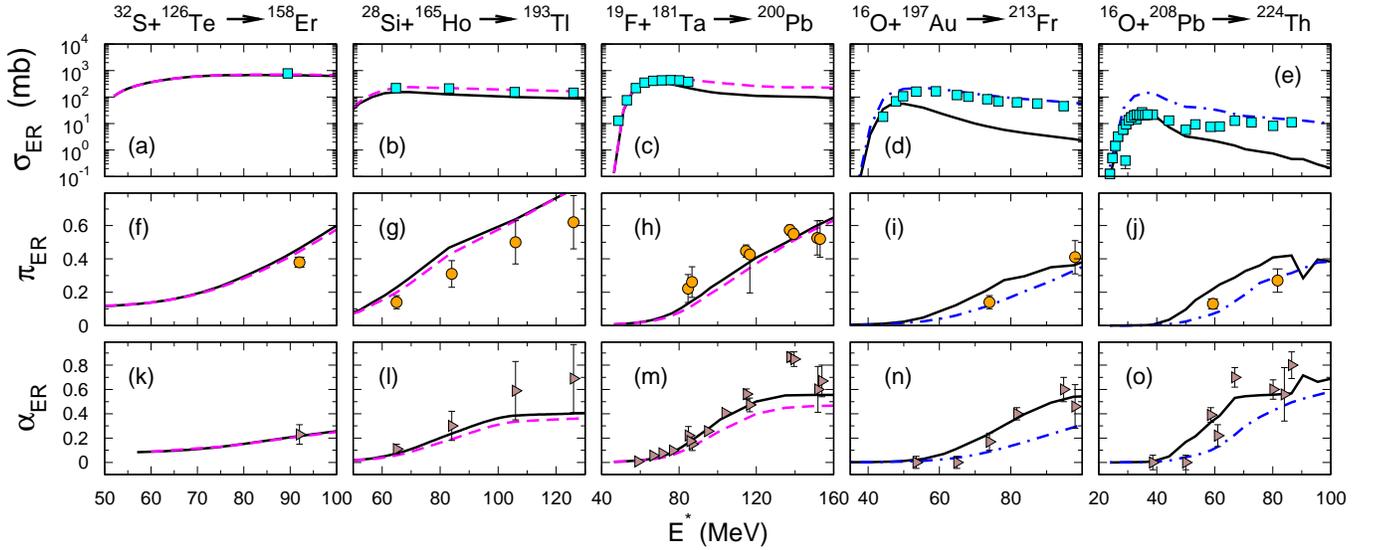}
\caption{\label{FIGURE7} Comparison of measured $\sigma_{\textrm{ER}}$, 
$\pi_{\textrm{ER}}$ and $\alpha_{\textrm{ER}}$ with SM predictions for the reactions:
$^{32}$S+$^{126}$Te \cite{Nadotchy2010}, $^{28}$Si+$^{165}$Ho
\cite{Fineman1994}, $^{19}$F+$^{181}$Ta
\cite{Hinde1986,Caraley2000,Fabris1994}, $^{16}$O+$^{197}$Au
\cite{Brinkmann1994,Fineman1994} and $^{16}$O+$^{208}$Pb
\cite{Fineman1994,Brinkmann1994,Morton1995}. Continuous (black) lines indicate the SM
predictions including shell correction in both $B_{\textrm{f}}$ and LD, CELD
and $K$-orientation effects. Dashed (magenta) and dash-dotted (blue) lines show
results with all the aforementioned effects and $\beta = 2 \times 10^{21}$
s$^{-1}$ and $\beta = 3 \times 10^{21}$ s$^{-1}$, respectvely.}
\end{center}
\end{figure*}
\begin{figure*}[ht!]
\begin{center}
\includegraphics[width=\textwidth,trim=0.0cm 0.0cm 0.0cm 0.0cm,clip]{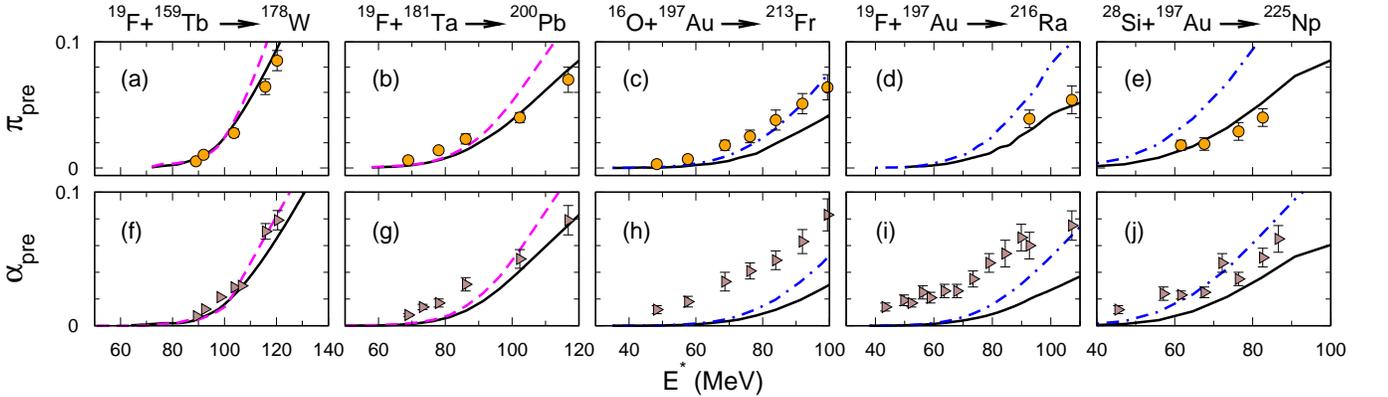}
\caption{\label{FIGURE8} Comparison of measured $\pi_{\textrm{pre}}$ and
$\alpha_{\textrm{pre}}$ with SM predictions for the reactions:
$^{19}$F+$^{159}$Tb \cite{Ikezoe1994}, $^{19}$F+$^{181}$Ta \cite{Ikezoe1992},
$^{16}$O+$^{197}$Au \cite{Ikezoe1992},
$^{19}$F+$^{197}$Au \cite{Ikezoe1992}, and $^{28}$Si+$^{197}$Au
\cite{Ikezoe1992}. Continuous (black) lines indicate the SM predictions
including shell correction in both $B_{\textrm{f}}$ and LD, CELD and
$K$-orientation effects. Dashed (magenta) and dash-dotted (blue) lines show
results with all the aforementioned effects and
$\beta = 2 \times 10^{21}$ s$^{-1}$ and $\beta = 3 \times 10^{21}$ s$^{-1}$,
respectvely.}
\end{center}
\end{figure*}

Numerous studies in the past have established that a slowing down of the
fission process, in comparison to that given by the Bohr-Wheeler fission width
is required in order to reproduce measured $\nu_{\textrm{pre}}$ [see e.g.
\cite{Lestone2009}]. In such cases, the available phase space at saddle (as in
Bohr-Wheeler theory) alone does not determine the fission rate, but the
dynamical evolution of the nuclear shape from ground state to the scission
point past the saddle point is to be taken into account. The process closely
resembles to the dynamics of a Brownian particle in a heat bath placed in a
potential pocket. The escape rate across the potential barrier or the fission
rate was obtained by Kramers \cite{Kramers1940} many years ago. A reduction in
fission width is obtained from the dissipative stochastic dynamical model of
fission developed by Kramers where the fission width is given as
\cite{Kramers1940}
\begin{figure*}[ht!]
\begin{center}
\includegraphics[width=\textwidth,trim=0.0cm 0.0cm 0.0cm 0.0cm,clip]{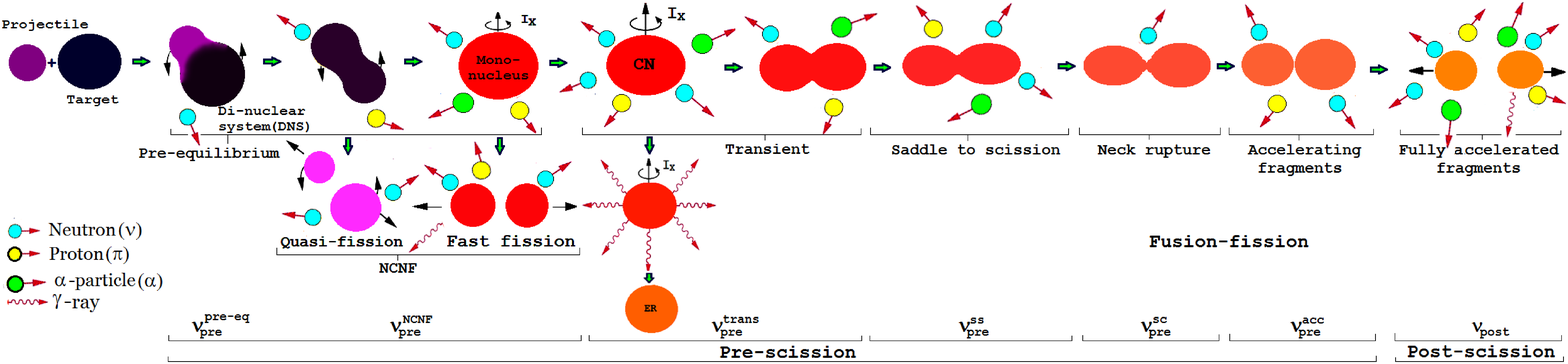}
\caption{\label{FIGURE9} Cartoon showing emission of neutrons at different
stages of heavy ion-induced fusion-fission reactions. See text for details.}
\end{center}
\end{figure*}
\begin{widetext}
\begin{equation}
\label{Kram}
\Gamma_{\textrm{f}}^{\textrm{Kram}} (E^{*}, \ell, K) =
\Gamma_{\textrm{f}}^{\textrm{BW}} (E^{*}, \ell, K)
\left\{ \sqrt{1+\left(\frac{\beta}{2\omega_{s}}\right)^{2}}-\frac{\beta}{2\omega_{s}} \right\} ,
\end{equation}
\end{widetext}
\noindent 
where $\beta$ is the reduced dissipation coefficient (ratio of dissipation
coefficient to inertia) and $\omega_{\textrm{s}}$ is the local frequency of a
harmonic oscillator potential which approximates the nuclear potential at the
saddle configuration and depends on the spin of the CN \cite{JhilamPRC78}.
$\Gamma_{\textrm{f}}^{\textrm{BW}} (E^{*}, \ell, K)$ is the Bohr-Wheeler
fission width obtained after incorporating shell corrected level densities,
CELD and $K$-orientation effect. Though nuclear dissipation has been a subject
of considerable amount of theoretical research, its precise nature and
magnitude is yet to be established \cite{Blocki1986,Hofmann1997,Mukho1997}. On
the other hand, extraction of dissipation coefficient from fitting of
experimental data is model dependent to some extent. We, therefore, choose to
treat $\beta$ as an adjustable parameter to fit experimental data in the
present work.

In a stochastic dynamical model of fission, a certain time interval elapses
before the fission rate reaches its stationary value as given by Eq. \ref{Kram}
\cite{Grange1983}. A parametrized form of time-dependent fission width is given
as \cite{BhattPRC33},
\begin{equation}
\Gamma_{\textrm{f}}^{\textrm{Kram}}(E^{*}, \ell, K, t) =
\Gamma_{\textrm{f}}^{\textrm{Kram}}(E^{*}, \ell, K) 
\left\{1-\textit{e}^{-\frac{2.3t}{\tau_{\textrm{f}}}}\right\}
\end{equation}
\noindent
where
\begin{equation}
\tau_{\textrm{f}} = \dfrac{\beta}{2\omega_{\textrm{g}}^{2}}\ln\left(\dfrac{10B_{\textrm{f}} (\ell)}{T}\right)
\end{equation}
\noindent
is the transient time period while the potential near the ground state is 
approximated by a harmonic oscillator potential of frequency
$\omega_{\textrm{g}}$.

The fission widths given by Bohr and Wheeler or Kramers are obtained under the
assumption that fission occurs when the CN shape crosses the saddle point
deformation. The number of neutrons evaporated prior to fission obtained in the
SM calculation therefore refers to those neutrons emitted till the CN reaches
the saddle point deformation. The experimentally determined
$\nu_{\textrm{pre}}$, however, includes all neutrons emitted till the CN splits
into two FFs. Thus neutrons emitted during saddle-to-scission transition of the
CN are also included in experimental $\nu_{\textrm{pre}}$. The
saddle-to-scission neutron multiplicity is obtained in the present SM
calculation by using the saddle-to-scission time interval which is given as
\cite{HofmannPLB122}
\begin{equation}
\label{SadScisTime}
\tau_{\textrm{ss}} = \tau_{\textrm{ss}}^{0} \left\{\sqrt{1+\left(\frac{\beta}
{2\omega_{\textrm{s}}}\right)^{2}}+\frac{\beta}{2\omega_{\textrm{s}}}\right\} ,
\end{equation}
\noindent
where $\tau_{\textrm{ss}}^{0}$ is the saddle-to-scission transit time without
any dissipation \cite{HofmannPLB122,Grange1983}.

The above features are incorporated in an SM code \textsc{vecstat}
\cite{TathaPLB2018}. Excitation functions of fission and ER and multiplicities
of evaporated light particles are calculated for a number of fusion-fission
reactions.

\subsection{Results}
The calculated $\sigma_{\textrm{ER}}$, $\sigma_{\textrm{fiss}}$ and
$\nu_{\textrm{pre}}$ ($\nu_{\textrm{ER}}$, in a few cases) for several
fusion-fission reactions are shown in Figs. \ref{FIGURE3} \textendash
\ref{FIGURE6} in increasing order of $A_{\textrm{CN}}$. The available
experimental values are also shown in the plots. The CN formed in the above
reactions range from that of a low fissility ($\chi_{\textrm{CN}} = 0.600$ for
$^{156}$Er) to a high one ($\chi_{\textrm{CN}} = 0.826$ for $^{248}$Cf).
Consequently, the dominating reaction products also change from ERs to FFs
across the systems considered here.

Two sets of SM results are displayed in Figs. \ref{FIGURE3} \textendash
\ref{FIGURE6}, where one set includes shell correction applied to both
$B_{\textrm{f}}(\ell)$ and level density and also CELD and $K$-orientation
effects but without any dissipation while the other set includes dissipation in
addition to the above-mentioned effects. It is observed that, in general, SM
results without dissipation overestimate $\sigma_{\textrm{fiss}}$ but
underestimate $\sigma_{\textrm{ER}}$ and $\nu_{\textrm{pre}}$. A value for
$\beta$ is next chosen to fit the experimental data. Since the effect of
dissipation in fission width in Eq. \ref{Kram} is obtained by considering
fission dynamics in the pre-saddle region \cite{Kramers1940}, $\beta$ in Eq.
\ref{Kram} also corresponds to the reduced dissipation coefficient in the same
region. Further, the most unambiguous signature of CN formation and subsequent
decay is the ER cross section. We, therefore, adjust the strength of $\beta$ in
order to fit the ER excitation function. It may, however, be mentioned that
$\beta$ in Eq. \ref{SadScisTime} represents the reduced dissipation coefficient
in the saddle-to-scission region. Though the strength of $\beta$ in the
pre-saddle and the post-saddle regimes need not be the same, we use the same
value for both in the present work.

It is observed that $\beta = 2 \times 10^{21}$ s$^{-1}$ gives reasonable fit to
ER and also to fission excitation functions up to $A_{\textrm{CN}} \sim 200$,
though $\nu_{\textrm{pre}}$ are under-predicted in certain cases (Figs.
\ref{FIGURE3} and \ref{FIGURE4}). A higher value of $\beta = 3 \times 10^{21}$
s$^{-1}$ is found to be necessary for CN between $^{210}$Po and $^{216}$Ra
(Figs. \ref{FIGURE5} and \ref{FIGURE6}) to fit the ER excitation functions.
Here too, the fission excitation functions are well reproduced but $\beta$ is
found to be inadequate for $\nu_{\textrm{pre}}$. For highly fissile CN
\textit{i.e.} $^{239}$Np and $^{248}$Cf, formed in light projectile-induced
reactions, the SM predictions for $\sigma_{\textrm{ER}}$ are very small and no
ER data are available (Fig. \ref{FIGURE6}). Since the calculated fission
excitation functions are insensitive to the strength of $\beta$, the same is
obtained by fitting the $\nu_{\textrm{pre}}$ for the above two CN. Good fits to
$\nu_{\textrm{pre}}$ excitation functions are obtained with
$\beta = 4 \times 10^{21}$ s$^{-1}$.

We next consider charged particle multiplicities in coincidence with either ERs
or FFs in fusion-fission reactions. SM predictions for proton and
$\alpha$-particle multiplicities in the ER channels for a number of reactions
are shown in Fig. \ref{FIGURE7} along with the experimental data. The ER
excitation functions are also given in this figure. The dissipation strength in
SM calculation is adjusted to fit the ER excitation functions and no other
parameter is tuned to fit the charged particle multiplicities. The fission
probability of CN has no direct effect on particle multiplicities from ERs
except influencing the ER angular momentum distribution and consequently, the
excitation energy available for evaporation. We, therefore, find that the
multiplicity distributions obtained with or without dissipation are very close
to each other for the systems $^{32}$S+$^{126}$Te, $^{28}$Si+$^{165}$Ho and
$^{19}$F+$^{181}$Ta since the ER excitation functions of the respective systems
obtained with or without dissipation are also close. On the other hand, charged
particle multiplicities for the systems $^{16}$O+$^{197}$Au and
$^{16}$O+$^{208}$Pb are significantly reduced and $\sigma_{\textrm{ER}}$ are
substantially increased when dissipation is included in the SM calculations.
This is a consequence of lowering of the average excitation energy of the
ERs since they carry larger angular momentum in SM calculations with
dissipation. We further observe that SM predictions match experimental data
reasonably well for both proton and $\alpha$-particle multiplicities from the
reactions $^{32}$S+$^{126}$Te, $^{28}$Si+$^{165}$Ho and $^{19}$F+$^{181}$Ta.
However, while close agreement with experimental data is observed for proton
multiplicity, the $\alpha$-multiplicity is underestimated to some extent for
the systems $^{16}$O+$^{197}$Au and $^{16}$O+$^{208}$Pb when dissipation is
included in SM calculation. It is also found in Fig. \ref{FIGURE7} that the SM
prediction with $\beta = 3 \times 10^{21}$ s$^{-1}$ overestimates
$\sigma_{\textrm{ER}}$ at lower excitation energies for the system
$^{16}$O+$^{208}$Pb though it fits the ER excitation function at higher
excitation energies. These are some of the issues which require further
investigation in future works.

Fig. \ref{FIGURE8} shows the pre-scission proton and $\alpha$-particle
multiplicities for a number of systems. Both experimental values and SM
predictions for charged particle multiplicities are given in this figure. SM
results are obtained both with and without dissipation. The dissipation
strength for a CN is taken to be the same as that used to fit the ER excitation
functions in the same mass region. The SM predictions qualitatively follow the
trend of experimental data. Similar to $\nu_{\textrm{pre}}$, SM calculation
yields larger values of multiplicities when dissipation is considered.
Quantitative comparison of SM predictions with experimental data, however, does
not show any definite pattern. While SM calculation with dissipation gives good
fit to experimental $\pi_{\textrm{pre}}$ for the $^{16}$O+$^{197}$Au and
experimental $\alpha_{\textrm{pre}}$ for the $^{28}$Si+$^{197}$Au, SM results
obtained without dissipation fit experimental $\pi_{\textrm{pre}}$ reasonably
well for $^{19}$F+$^{181}$Ta, $^{19}$F+$^{197}$Au and $^{28}$Si+$^{197}$Au
systems. Moreover, inclusion of dissipation though improves SM results for
$\alpha_{\textrm{pre}}$ for the $^{16}$O+$^{197}$Au and $^{19}$F+$^{197}$Au,
they still considerably underestimate the experimental values. It may be
remarked here that some part of evaporation from a CN undergoing fission can
take place when it is deformed, particularly during the saddle-to-scission
transition. The effect of deformation on decay widths is expected to be higher
for charged particles than neutrons because of the Coulomb potential
\cite{Lestone1993} which is not included in the present calculations. This may
account for the deviations of SM predictions from the experimental data in
certain cases. Accurate charged particles multiplicity data for more systems in
both ER and fission channels will help to improve the SM further.

Thus, SM analysis of a large number of systems covering a broad range of
$A_{\textrm{CN}}$ and $\chi_{\textrm{CN}}$ show that the ER and fission
excitation functions can be well reproduced when a $\beta$ of strength in the
range 2\textendash4 $\times 10^{21}$ s$^{-1}$ is included along with shell
effects both in $B_{\textrm{f}}$ and LD, CELD and $K$-orientation effect.
Dissipation strength of the above magnitude in fission dynamics has also been
found necessary in earlier works \cite{Lestone2009,Dioszegi2000,Frobrich1998}.
Though SM predictions for $\nu_{\textrm{pre}}$ are found to be satisfactory for
some systems, they tend to underestimate it for the others. The limitations of
the SM in predicting pre-scission particle multiplicities are discussed in the
next Section.

\section{Discussion}
\label{Dis}
In general, the neutron (or any other light charged particle) multiplicity
in a heavy ion induced fusion-fission reaction comprises of neutrons emitted at
the following stages (see Fig. \ref{FIGURE9}) of the reaction:
\begin{itemize}
\item pre-equilibrium composite system
	($\nu_{\textrm{pre}}^{\textrm{pre-eq}}$),
\item fragments originating from NCNF of the composite system
	($\nu_{\textrm{pre}}^{\textrm{NCNF}}$), 
\item transient: ground state of CN to saddle
	($\nu_{\textrm{pre}}^{\textrm{trans}}$),
\item saddle-to-scission ($\nu_{\textrm{pre}}^{\textrm{ss}}$), 
\item neck rupture ($\nu_{\textrm{pre}}^{\textrm{sc}}$),
\item acceleration phase of fission fragments
	($\nu_{\textrm{pre}}^{\textrm{acc}}$),
\item post-scission: fully accelerated fission fragments
	($\nu_{\textrm{post}}$). 
\end{itemize}
 
The pre-scission component of the experimental neutron multiplicity
($\nu_{\textrm{pre}}^{\textrm{exp}}$), which is compared with SM predictions
would thus be given as, 
\begin{equation}
\label{NuPre}
\nu_{\textrm{pre}}^{\textrm{exp}} = \nu_{\textrm{pre}}^{\textrm{pre-eq}}+
\nu_{\textrm{pre}}^{\textrm{NCNF}}+\nu_{\textrm{pre}}^{\textrm{trans}}
+\nu_{\textrm{pre}}^{\textrm{ss}}+\nu_{\textrm{pre}}^{\textrm{sc}}+
\nu_{\textrm{pre}}^{\textrm{acc}}.
\end{equation}

The SM, however, includes only $\nu_{\textrm{pre}}^{\textrm{trans}}$ and
$\nu_{\textrm{pre}}^{\textrm{ss}}$ in the calculated $\nu_{\textrm{pre}}$
\cite{HofmannPLB122}. Evidently, good agreement between calculated and
experimental multiplicities is expected when the contributions of other terms
in Eq. \ref{NuPre}, which are not included in the SM, are small. Further, the
relative magnitudes of $\nu_{\textrm{pre}}^{\textrm{trans}}$ and
$\nu_{\textrm{pre}}^{\textrm{ss}}$ depend upon the angular momentum of the CN
since the fission barrier decreases with increasing angular momentum and
consequently $\nu_{\textrm{pre}}^{\textrm{trans}}$ decreases. The
saddle-to-scission interval also increases with increasing angular momentum.
Thus, the relative contribution of $\nu_{\textrm{pre}}^{\textrm{ss}}$ increases
with increasing angular momentum of the CN. The CN shape is strongly deformed
in the saddle-to-scission region though its effect on particle widths has not
been taken into account in the present calculation. Further,
$\nu_{\textrm{pre}}^{\textrm{ss}}$ is obtained from an empirical formulation of
saddle-to-scission transit time, given by Eq. \ref{SadScisTime}. Consequently,
the calculated $\nu_{\textrm{pre}}^{\textrm{ss}}$ can be uncertain to certain
extent. It is, therefore, possible that discrepancy between experimental values
and SM predictions of $\nu_{\textrm{pre}}$ increases with projectile mass for
the same CN. Such a trend is observed in general in Figs. \ref{FIGURE3}
\textendash \ref{FIGURE6} with the exception of $^{4}$He+$^{188}$Os,$^{206}$Pb.
The exception is possibly due to emission in the pre-equilibrium stage
\cite{Sarantites1978,Singh1992,Chakravarty1992,Knoche1995,Singh2006} which will
be discussed shortly. It may also be mentioned here that a stronger dissipation
in the saddle-to-scission region is expected to give a better fit to the
experimental $\nu_{\textrm{pre}}$. Such a shape-dependent dissipation has been
reported earlier from phenomenological studies \cite{Dioszegi2000,Frobrich1998}
and also from theoretical considerations \cite{Pal1998}.

One characteristic of statistical decay of CN is the expectation that the
$\gamma$-ray multiplicity ($\langle M_{\textit{x}} \rangle$) for a channel
corresponding to the emission of \textit{x}-neutrons should increase with
decreasing \textit{x} as was observed for $^{20}$Ne+$^{150}$Nd, since
evaporation of fewer neutrons would leave more energy for $\gamma$-ray emission
\cite{SarantitesPRC1978}. However, this was not the case for
$^{12}$C+$^{158}$Gd, where, $\langle M_{\textit{x}} \rangle$ remained
essentially constant for small \textit{x}. This saturation effect observed in
the $\gamma$-ray multiplicity of $^{12}$C+$^{158}$Gd, relative to the same of
$^{20}$Ne+$^{150}$Nd at the same CN excitation energy (and slightly different
angular momenta), was said to be a clear signature of pre-equilibrium
emissions of neutrons
\cite{Blann1974,SarantitesPRC1978,Sarantites1978,Westerberg1978}.
The disagreement of theory and measurement for $^{4}$He+$^{188}$Os,$^{206}$Pb
may also be attributed to the emissions in the pre-equilibrium stage
\cite{Sarantites1978,Singh1992,Chakravarty1992,Knoche1995,Singh2006}.

For systems with higher mass-symmetry, neutrons
($\nu_{\textrm{pre}}^{\textrm{exp}}$) can also be emitted in the 
comparatively longer formation stage
($\nu_{\textrm{pre}}^{\textrm{pre-eq}}+\nu_{\textrm{pre}}^{\textrm{NCNF}}$)
of the CN \cite{Saxena1994,Vigdor1982,Cabrera2003} and / or during neck rupture
($\nu_{\textrm{pre}}^{\textrm{sc}}$)
\cite{Bowman1962,Loveland1967,Halpern1971,Franklyn1978,Brosa1983,Hwang1999,Kornilov2007,Carjan2010,Capote2016} 
and / or from the acceleration phase of fission fragments
($\nu_{\textrm{pre}}^{\textrm{acc}}$)
\cite{Eismont1965,Cheifetz1970,Skarsvag1973,Hinde1984}, which are not included
in the present work. These are the most probable reasons for the mismatch in
the experimental and the calculated $\nu_{\textrm{pre}}$ for the symmetric
combinations. Further, the $\nu_{\textrm{pre}}^{\textrm{exp}}$ were found to be
higher than those extracted from fission chance distributions
\cite{Golda2013,Mahata2017}. This indicates a significant post-saddle
contribution in the $\nu_{\textrm{pre}}^{\textrm{exp}}$. It must be mentioned
here that unlike scission neutrons, protons, due to their Coulomb repulsion,
have less presence in the neck region \cite{Carjan2015}.

ER is the clearest signature of fusion and the particle multiplicities
extracted in coincidence with ERs are not affected by the emissions from the
later part i.e. fission (saddle-to-scission dynamics). Therefore, ERs from
highly asymmetric systems can serve as the benchmark for input parameters of
the SM which fit the experimental data. For example, particle multiplicity data
(in coincidence with ERs as well as FFs) of the asymmetric reaction
$^{19}$F+$^{181}$Ta (Fig. \ref{FIGURE7} and Fig. \ref{FIGURE8}) are reproduced
quite well with the present SM calculations. This indicates that deformation
effects in saddle-to-scission region and other near-scission and post-scission
contributions are not severe for this system. However, this is not the case for
few other systems $\textit{e.g.}$ $^{19}$F,$^{28}$Si+$^{197}$Au and the
discrepancy possibly arises from the aforementioned processes. Moreover,
dynamical deformation may cause the observed deviation of the predicted
particle multiplicities from the measured ones \cite{Ye2010,Neeraj2017}.

It has been noticed earlier that while a standard set of parameters in an SM
could reproduce the gross features of $\alpha$-particle spectra from the ER
channel, they failed in the fission channel \cite{Vardaci1998}. A reduction in
the emission barrier was necessary for a satisfactory reproduction of the
latter. This was attributed to the large deformation of the fissioning nuclei
with respect to the ER channel 
\cite{Ikezoe1990,Lestone1993,Vardaci1998}. With increasing deformation, the
binding energies of the charged particles (proton, $\alpha$-particles etc.)
increase whereas the effective emission barrier heights decrease. While the
former lowers the emission rate, the latter enhances it
\cite{BeckermanPRL1979,Blann1980}. Altogether, accounting for the effect of
deformation energy, particle transmission co-efficients and particle binding
energies on the emission rates, leads to the suppression of the
$\pi_{\textrm{pre}}$ and $\alpha_{\textrm{pre}}$ compared to
$\nu_{\textrm{pre}}$ \cite{Lestone1991}. Analysis of the kinetic energy spectra
of the emitted light charged particles and / or an estimation of the SM
branching ratios by analysing measured $\nu_{\textrm{ER}}$ can shed light on
the origin of the overestimation of the calculated multiplicities of the
symmetric reactions by the SM \cite{Wile1995,Nitto2011,Kalandarov2012}.
Aleshin \cite{Aleshin1993} presented a semi-classical description for light
particle emission from a composite system with a time-dependent shape. A
satisfactory description of the charged particle multiplicities was achieved by
taking the mean value of the particle separation energies of the two nuclei
making the composite system rather than that of the mononucleus (which 
overestimated multiplicities) while calculating the decay widths. These aspects
need further investigation.

\section{Conclusion}
\label{Conc}
We have carried out a systematic analysis of available $\sigma_{\textrm{ER}}$,
$\sigma_{\textrm{fiss}}$, $\nu_{\textrm{pre}}$, $\pi_{\textrm{pre}}$,
$\alpha_{\textrm{pre}}$ (also $\nu_{\textrm{ER}}$, $\pi_{\textrm{ER}}$ and
$\alpha_{\textrm{ER}}$) data covering $A_{\textrm{CN}}=158$ \textendash 248.
Shell effect in LD and shell correction in $B_{\textrm{f}}$, effect of
$K_{\textrm{or}}$, CELD and dissipation in fission have been considered in the
SM. Parameters of the model have not been tuned to reproduce observables from
specific reaction(s) except for $\beta$, strength of the reduced dissipation
coefficient, which has been treated as the only adjustable parameter in the
calculation. The model is able to reproduce
$\sigma_{\textrm{ER}}$ and $\sigma_{\textrm{fiss}}$ simultaneously for all the
reactions considered in this work. Experimental $\nu_{\textrm{pre}}$ are
underestimated by the calculation in some cases. An increasingly higher
strength of $\beta$ (2 \textendash 4 $\times 10^{21}$ s$^{-1}$) is required to
reproduce the data with increasing $A_{\textrm{CN}}$. Experimental proton and
$\alpha$-particle multiplicities, measured in coincidence with both ERs and
FFs, are in qualitative agreement with model predictions. The present
investigation, thus, mitigates the existing inconsistencies in modelling
statistical decay of the fissile CN to a large extent. Contradictory
requirements of fission enhancement, by scaling down $B_{\textrm{f}}$, to
reproduce $\sigma_{\textrm{ER}}$ or $\sigma_{\textrm{fiss}}$ and fission
suppression, by introducing dissipation in the fission channel, to reproduce
$\nu_{\textrm{pre}}$ for similar reactions, is no longer called for. There are
scopes for further refinement of the model, particularly in the domains of
dynamical effects in the post-saddle and near-scission regions for neutrons and
deformation effects on the charged particle emission widths in the post-saddle
region, as is evident from the mismatch between measured and calculated light
particle multiplicities in a few cases.

\vspace{7.5mm}
\section{Acknowledgement}
T.B. acknowledges IUAC, New Delhi, for financial support in the form of a
fellowship (IUAC/FS-1027/5876). S.P. acknowledges the kind hospitality provided
by IUAC through Visiting Associateship during the course of this work.

\newpage
\vspace{10mm}


\noindent

\begin{thebibliography}{120}
\bibitem{Bohr1936} N. Bohr, Nature (London) \textbf{137}, 344 (1936).
\bibitem{TathaPLB2018} Tathagata Banerjee, S. Nath and Santanu Pal, Phys. Lett.
B \textbf{776}, 163 (2018).
\bibitem{Schmitt2014} C. Schmitt, K. Mazurek, P. N. Nadtochy, 
Phys. Lett. B \textbf{737}, 289 (2014).
\bibitem{Delag1977} H. Delagrange, A. Fleury, and J. M. Alexander, 
Phys. Rev. C \textbf{16}, 706 (1977).
\bibitem{Videbaek1977} F. Videbæk, R. B. Goldstein, L. Grodzins, S. G. Steadman,
T. A. Belote, and J. D. Garrett, Phys. Rev. C \textbf{15}, 954 (1997).
\bibitem{Ward1983} D. Ward, R. J. Charity, D. J. Hinde, J. R. Leigh, J. O.
Newton, Nucl. Phys. \textbf{A403}, 189 (1983).
\bibitem{LestonePRL1993} J. P. Lestone, Phys. Rev. Lett. \textbf{70}, 2245
(1993).
\bibitem{Fineman1994} B. J. Fineman, K.-T. Brinkmann, A. L. Caraley, N. Gan,
R. L. McGrath, and J. Velkovska, Phys. Rev. C \textbf{50}, 1991 (1994).
\bibitem{Charity2010} R. J. Charity, Phys. Rev. C \textbf{82}, 014610 (2010).
\bibitem{Nitto2011} A. Di Nitto, E. Vardaci, A. Brondi, G. La Rana, R. Moro,
P.N. Nadtochy, M. Trotta, A. Ordine, A. Boiano, M. Cinausero, et al., Eur. Phys.
J. A \textbf{47}, 83 (2011).
\bibitem{Mancusi2010} Davide Mancusi, Robert J. Charity, and Joseph Cugnon, 
Phys. Rev. C \textbf{82}, 044610 (2010).
\bibitem{McCalla2008} S. G. McCalla and J. P. Lestone, Phys. Rev. Lett.
\textbf{101}, 032702 (2008).
\bibitem{Lestone2009} J. P. Lestone and S. G. McCalla, Phys. Rev. C \textbf{79},
044611 (2009).
\bibitem{Brinkmann1994} K.-T. Brinkmann, A. L. Caraley, B. J. Fineman, N. Gan,
J. Velkovska, and R. L. McGrath, Phys. Rev. C \textbf{50}, 309 (1994).
\bibitem{Sagaidak2009} R. N. Sagaidak and A. N. Andreyev, Phys. Rev. C
\textbf{79}, 054613 (2009).
\bibitem{Varinderjit2014} Varinderjit Singh, B. R. Behera, Maninder Kaur, A.
Kumar, K. P. Singh, N. Madhavan, S. Nath, J. Gehlot, G. Mohanto, A. Jhingan,
et al., Phys. Rev. C \textbf{89}, 024609 (2014).
\bibitem{Hinde1986} D. J. Hinde, R. J. Charity, G. S. Foote, J. R. Leigh, J. O.
Newton, S. Ogaza, A. Chattejee, Nucl. Phys. \textbf{A452}, 550 (1986).
\bibitem{Newton1988} J. O. Newton, D. J. Hinde, R. J. Charity, J. R. Leigh,
J. J. M. Bokhorst, A. Chatterjee, G. S. Foote, and S. Ogaza, Nucl. Phys.
\textbf{A483}, 126 (1988).
\bibitem{Vardaci2010} E. Vardaci, A. Di Nitto, A. Brondi, G. La Rana, R. Moro,
P.N. Nadotchy, M. Trotta, A. Ordine, A. Boiano, M. Cinausero, et al., Eur. Phys.
J. A \textbf{43}, 127 (2010).
\bibitem{Hilscher1992} D. Hilscher and H. Rossner, Ann. Phys. Fr. \textbf{17},
471 (1992).
\bibitem{Paul1994} P. Paul and M. Thoennessen, Annu. Rev. Nucl. Part. Sci.
\textbf{44}, 55 (1994).
\bibitem{Dioszegi2000} I. Di\'{o}szegi, N. P. Shaw, I. Mazumdar, A.
Hatzikoutelis, and P. Paul, Phys. Rev. C \textbf{61}, 024613 (2000).
\bibitem{Frobrich1998} P. Fr\"{o}brich and I. I. Gontchar, Phys. Rep.
\textbf{292}, 131 (1998).
\bibitem{Shaw2000} N. P. Shaw, I. Di\'{o}szegi, I. Mazumdar, A. Buda, C. R.
Morton, J. Velkovska, J. R. Beene, D. W. Stracener, R. L. Varner, M.
Thoennessen, and P. Paul, Phys. Rev. C \textbf{61}, 044612 (2000).
\bibitem{Laveen2015} P. V. Laveen, E. Prasad, N. Madhavan, S. Pal, J. Sadhukhan,
S. Nath, J. Gehlot, A. Jhingan, K. M. Varier, R. G. Thomas, A. M. Vinodkumar,
A. Shamlath, T. Varughese, P. Sugathan, B. R. S. Babu, S. Appannababu, K. S.
Golda, B. R. Behera, Varinderjit Singh, Rohit Sandal, A. Saxena, B. V. John,
and S. Kailas, J. Phys. G: Nucl. Part. Phys. \textbf{42}, 095105 (2015).
\bibitem{Mohanto2012} Gayatri Mohanto, N. Madhavan, S. Nath, Jhilam Sadhukhan,
J. Gehlot, I. Mazumdar, M. B. Naik, E. Prasad, Ish Mukul, T. Varughese,
A. Jhingan, R. K. Bhowmik, A. K. Sinha, D. A. Gothe, P. B. Chavan, Santanu Pal,
V. S. Ramamurthy, and A. Roy, Nucl. Phys. A \textbf{890\textendash891}, 62
(2012).
\bibitem{Ramachandran2006} K. Ramachandran, A. Chatterjee, A. Navin, K. Mahata,
A. Shrivastava, V. Tripathi, S. Kailas, V. Nanal, R. G. Pillay, A. Saxena, R. G.
Thomas, Suresh Kumar, and P. K. Sahu, Phys. Rev C \textbf{73}, 064609 (2006).
\bibitem{Tatha2015} Tathagata Banerjee, S. Nath, and Santanu Pal, Phys. Rev. C
\textbf{91}, 034619 (2015).
\bibitem{Janssens1986} R. V. F. Janssens, R. Holzmann, W. Henning, T. L. Khoo,
K. T. Lesko, G. S. F. Stephans, D. C. Radford, A. M. Van Den Berg, Phys. Lett.
\textbf{B181}, 16 (1986).
\bibitem{Wolfs1989} F. L. H. Wolfs, R. V. F. Janssens, R. Holzmann, T. L. Khoo,
W. C. Ma, and S. J. Sanders, Phys. Rev., C \textbf{39}, 865 (1989).
\bibitem{Zebelman1973} A. M. Zebelman and J. M. Miller, Phys. Rev. Lett.
\textbf{30}, 27 (1973).
\bibitem{GavronPRL1981} A. Gavron, J. R. Beene, B. Cheynis, R. L. Ferguson,
F. E. Obenshain, F. Plasil, G. R. Young, G. A. Petitt,
M. J\"{a}\"{a}skel\"{a}inen, D. G. Sarantites, and C. F. Maguire, 
Phys. Rev. Lett. \textbf{47}, 1255 (1981); \textbf{48}, 835(E) (1982).
\bibitem{Leigh1995} J. R. Leigh, M. Dasgupta, D. J. Hinde, J. C. Mein, C. R.
Morton, R. C. Lemmon, J. P. Lestone, J. O. Newton, H. Timmers, J. X. Wei, and
N. Rowley, Phys. Rev. C \textbf{52}, 3151 (1995).
\bibitem{HindeRapid1988} D. J. Hinde, H. Ogata, M. Tanaka, T. Shimoda, N.
Takahashi, A. Shinoharat, S. Wakamatsu, K. Katori and H. Okamura, Phys. Rev. C
\textbf{37}, 2923 (1988).
\bibitem{Hinde1992} D. J. Hinde, D. Hilscher, H. Rossner, B. Gebauer, M.
Lehmann, and M. Wilpert, Phys. Rev. C \textbf{45}, 1229 (1992).
\bibitem{Sarantites1976} D. G. Sarantites, J. H. Barker, M. L. Halbert, D. C.
Hensley, R. A. Dayras, E. Eichler, N. R. Johnson, and S. A. Gronemeyer,
Phys. Rev. C \textbf{14}, 2138 (1976).
\bibitem{Zebelman1974} A. M. Zebelman, L. Kowalski, J. Miller, K. Beg, Y. Eyal,
G. Jaffe, A. Kandil, and D. Logan, Phys. Rev. C \textbf{10}, 200 (1974).
\bibitem{Halbert1978} M. L. Halbert, R. A. Dayras, R. L. Ferguson, F. Plasil,
and D. G. Sarantites, Phys. Rev. C \textbf{17}, 155 (1978).
\bibitem{Navin2004} A. Navin, V. Tripathi, Y. Blumenfeld, V. Nanal, C. Simenel,
J. M. Casandjian, G. de France, R. Raabe, D. Bazin, A. Chatterjee, M. Dasgupta,
S. Kailas, R. C. Lemmon, K. Mahata, R. G. Pillay, E. C. Pollacco, K.
Ramachandran, M. Rejmund, A. Shrivastava, J. L. Sida, and E. Tryggestad, 
Phys. Rev. C \textbf{70}, 44601 (2004).
\bibitem{IgnatyukJNucl1975} A. V. Ignatyuk, M. G. Itkis, V. N. Okolovich, G. N.
Smirenkin and A. S. Tishin, ``Fission of pre-actinide nuclei Excitation
functions for the ($\alpha$,f) reaction", Yad. Fiz. \textbf{21(6)}, 1185 (1975);
Sov. J. Nucl. Phys., \textbf{21} 612 (1976).
\bibitem{Schmitt2003} R. P. Schmitt, T. Botting, G. G. Chubarian, K. L. Wolf,
B. J. Hurst, H. Jabs, M. Hamelin, A. Bacak, Yu. Ts. Oganessian, M. G. Itkis,
E. M. Kozulin, et al., Phys. of Atom. Nucl. \textbf{66}, 1163 (2003).
\bibitem{Tapan2016} Tapan Rajbongshi, K. Kalita, S. Nath, J. Gehlot, Tathagata
Banerjee, Ish Mukul, R. Dubey, N. Madhavan, C. J. Lin, A. Shamlath, P. V.
Laveen, M. Shareef, Neeraj Kumar, P. Jisha, and P. Sharma, Phys. Rev. C
\textbf{93}, 54622 (2016).
\bibitem{Devendra2009} Devendra Singh, Unnati, Pushpendra P. Singh, Abhishek
Yadav, Manoj Kumar Sharma, B. P. Singh, K. S. Golda, Rakesh Kumar, A. K. Sinha,
and R. Prasad, Phys. Rev. C \textbf{80}, 014601 (2009).
\bibitem{Ogihara1990} M. Ogihara, H. Fujiwara, S. C. Jeong, W. Galster, S. M.
Lee, Y. Nagashima, T. Mikumo, H. Ikezoe, K. Ideno, Y. Sugiyama, Y. Tomita, Z.
Phys. A \textbf{335}, 203 (1990).
\bibitem{Bivash2001} Bivash R. Behera, Subinit Roy, P. Basu, M. K. Sharan, S.
Jena, M. Satpathy, M. L. Chatterjee, S. K. Datta, Pramana - J. Phys.
\textbf{57}, 199 (2001).
\bibitem{Hardev2007} Hardev Singh, Ajay Kumar, Bivash R. Behera, I. M. Govil,
K. S. Golda, Pankaj Kumar, Akhil Jhingan, R. P. Singh, P. Sugathan, M. B.
Chatterjee, S. K. Datta, Ranjeet, Santanu Pal, and G. Viesti, Phys. Rev. C
\textbf{76}, 044610 (2007).
\bibitem{Shidling2006} P. D. Shidling, N. M. Badiger, S. Nath, R. Kumar,
A. Jhingan, R. P. Singh, P. Sugathan, S. Muralithar, N. Madhavan, A. K. Sinha,
Santanu Pal, et al., Phys. Rev. C \textbf{74}, 064603 (2006).
\bibitem{Leigh1988} J. R. Leigh, J. J. M. Bokhorst, D. J. Hinde and J. O.
Newton, J. Phys. G: Nucl. Phys. \textbf{14}, L55 (1988).
\bibitem{Bemis1986} C. E. Bemis Jr., T. C Awes, J. R. Beene, R. L. Ferguson,
H. J. Kim, F. K. McGowan, F. E. Obenshain, F. Plasil, P. Jacobs, Z. Frankel,
U. Smilansky, I. Tserruya, ORNL physics division progress report 6326 (1986).
\bibitem{Forster1987} J. S. Forster, I. V. Mitchell, J. U. Andersen, A. S.
Jensen, E. Laegsgaard, W. M. Gibson, K. Reichelt, Nucl. Phys. \textbf{A464}, 497
(1987).
\bibitem{Charity1986} R. J. Charity, J. R. Leigh, J. J. M. Bokhorst, A.
Chatterjee, G. S. Foote, D. J. Hinde, J. O. Newton, S. Ogaza, and D. Ward,
Nucl. Phys. \textbf{A457}, 441 (1986).
\bibitem{Hinde1982} D. J. Hinde, J. R. Leigh, J. O. Newton, W. Galster, and
S. Sie, Nucl. Phys. \textbf{A385} 109 (1982).
\bibitem{Gayatri2013} G. Mohanto, N. Madhavan, S. Nath, J. Gehlot, I. Mukul, 
A. Jhingan, T. Varughese, A. Roy, R. K. Bhowmik, I. Mazumdar, D. A. Gothe, 
P. B. Chavan, J. Sadhukhan, S. Pal, M. Kaur, V. Singh, A. K. Sinha, and V. S. 
Ramamurthy, Phys. Rev. C \textbf{88}, 034606 (2013).
\bibitem{Beyec1970} Y. Le Beyec et M. Lefort, Nucl. Phys. \textbf{A99}, 131
(1967).
\bibitem{Gadioli1969} E. Gadioli, I. Iori, N. Molho, and L. Zetta, Lett. Nuovo
Cimento \textbf{2}, 904 (1969).
\bibitem{Khodai1966} A. Khodai-Joopari, Ph.D. Thesis, Report UCRL-16489,
Berkeley (1966).
\bibitem{Ignatyuk1984} A. V. Ignatyuk, M. G. Itkis, I. A. Kamenev, S. I. Mulgin,
V. N. Okolovich, G. N. Smirenkin, Yad. Fiz. \textbf{40}, 625 (1984) [Sov. J.
Nucl. Phys. \textbf{40}, 400 (1984)].
\bibitem{Shigaev1973} O. E. Shigaev, V. S. Bychenkov, M. F. Lomanov, A. I.
Obukhov, N. A. Perfilov, G. G. Shimchuk, and R. M. Yakovlev, V. G. Khlopin
Radium Institute, St. Petersburg, Preprint RI-17 (1973); Yadernaya Fizika 17,
947 (1973) [Sov. J. Nucl. Phys. \textbf{17}, 496 (1974)].
\bibitem{Cheifetz1970} E. Cheifetz, Z. Fraenkel, J. Galin, M. Lefort, J. Péter,
and X. Tarrago, Phys. Rev. C \textbf{2}, 256 (1970).
\bibitem{Bimbot1969} Par R. Bimbot, H. Zaffrezic, Y. Le Beyec, M. Lefort et A.
Vigny-Simon, Le Journal de Physique, \textbf{30}, 513 (1969).
\bibitem{Shrivastava1999} A. Shrivastava, S. Kailas, A. Chatterjee, A. Navin,
A. M. Samant, P. Singh, S. Santra, K. Mahata, B. S. Tomar, and G. Pollarolo,
Phys. Rev. C \textbf{63}, 054602 (2001).
\bibitem{Plicht1983} J. Vander Plicht, H. C. Britt, M. M. Fowler, Z. Fraenkel,
A. Gavron, J. B. Wilhelmy, F. Plasil, T. C. Awes, and G. R. Young, Phys. Rev. C
\textbf{28}, 2022 (1983).
\bibitem{Golda2013} K. S. Golda, A. Saxena, V. K. Mittal, K. Mahata, P.
Sugathan, A. Jhingan, V. Singh, R. Sandal, S. Goyal, J. Gehlot, A. Dhal, B. R.
Behera, R. K. Bhowmik, S. Kailas, Nucl. Phys. \textbf{A913}, 157 (2013).
\bibitem{Hinde2002} D. J. Hinde, A. C. Berriman, R. D. Butt, M. Dasgupta,
I. I. Gontchar, C. R. Morton, A. Mukherjee, and J. O. Newton, J. Nucl.
Radiochem. Sci. \textbf{3}, 31 (2002).
\bibitem{Sagaidak2003} R. N. Sagaidak, G. N. Kniajeva, I. M. Itkis, M. G. Itkis, 
N. A. Kondratiev, E. M. Kozulin, I. V. Pokrovsky, A. I. Svirikhin, V. M.
Voskressensky, A. V. Yeremin, et al., Phys. Rev. C \textbf{68}, 014603 (2003).
\bibitem{Hardev2008} Hardev Singh, K. S. Golda, Santanu Pal, Ranjeet, Rohit
Sandal, Bivash R. Behera, Gulzar Singh, Akhil Jhingan, R. P. Singh, P. Sugathan, 
M. B. Chatterjee, S. K. Datta, Ajay Kumar, G. Viesti, and I. M. Govil,
Phys. Rev. C \textbf{78}, 024609 (2008).
\bibitem{Bate1964} George L. Bate and J. R. Huizenga, Phys. Rev. C \textbf{133},
B1471 (1964).
\bibitem{Kandil1976} A. T. Kandil, J. Inorg. Nucl. Chem. \textbf{38}, 37 (1976).
\bibitem{Strecker1990} M. Strecker, R. Wein, P. Plischke, and W. Scobel, 
Phys. Rev. C \textbf{41}, 2172 (1990).
\bibitem{Rubchenya2001} V. A. Rubchenya, W. H. Trzaska, D. N. Vakhtin, J.
\"{A}yst\"{o}, P. Dendooven, S. Hankonen, A. Jokinen, Z. Radivojevich, J. C.
Wang, I. D. Alkhazov, A. V. Evsenin, et al., Nucl. Instr. and Meth. A
\textbf{463}, 653 (2001).
\bibitem{Freies1975} H. Freiesleben, G. T. Rizzo, J. R. Huizenga, Phys. Rev. C
\textbf{12}, 42 (1975).
\bibitem{Hinde1989} D. J. Hinde, H. Ogata, M. Tanaka, T. Shimoda, N. Takahashi,
A. Shinohara, S. Wakamatsu, K. Katori, and H. Okamura, Phys. Rev. C \textbf{39},
2268 (1989).
\bibitem{Liu1996} Zuhua Liu, Huanqiao Zhang, Jincheng Xu, Yu Qiao, Xing Qian,
and Chengjian Lin, Phys. Rev. C \textbf{54}, 761 (1996).
\bibitem{Kailas1999} S. Kailas, D. M. Nadkarni, A. Chatterjee, A. Saxena, S. S.
Kapoor, R. Vandenbosch, J. P. Lestone, J. F. Liang, D. J. Prindle, A. A.
Sonzogni, and J. D. Bierman, Phys. Rev. C \textbf{59}, 2580 (1999).
\bibitem{Saxena1994} A. Saxena, A. Chatterjee, R. K. Choudhury, S. S. Kapoor,
and D. M. Nadkarni, Phys. Rev. C \textbf{49}, 932 (1994).
\bibitem{Back1985} B. B. Back, R. R. Betts, J. E. Gindler, B. D. Wilkins, S.
Saini, M. B. Tsang, C. K. Gelbke, W. G. Lunch, M. A. McMahan, and P. A. Baisden, 
Phys. Rev. C \textbf{32}, 195 (1985).
\bibitem{Vandenbosch1986} R. Vandenbosch, T. Murakami, C. -C. Sahm, D. D. Leach,
A. Ray, and M. J. Murphy, Phys. Rev. Lett. \textbf{56}, 1234 (1986).
\bibitem{Nadkarni1999} D. M. Nadkarni, A. Saxena, D. C. Biswas, R. K. Choudhury,
S. S. Kapoor, N. Majumdar, and P. Bhattacharya, Phys. Rev. C \textbf{59},
R580(R) (1999).
\bibitem{Nadotchy2010} P. N. Nadotchy, A. Brondi, A. Di Nitto, G. La Rana, R.
Moro, E. Vardaci, A. Ordine, A. Boiano, M. Cinausero, G. Prete, V. Rizzi, N.
Gelli and F. Lucarelli, EPJ Web Conf. \textbf{2}, 08003 (2010).
\bibitem{Ikezoe1994} H. Ikezoe, Y. Nagame, I. Nishinaka, Y. Sugiyama, Y. Tomita,
K. Ideno, S. Hamada, N. Shikazono, A. Iwamoto and T. Ohtsuki, Phys. Rev. C
\textbf{49}, 968 (1994).
\bibitem{Ikezoe1992} H. Ikezoe, N. Shikazono, Y. Nagame, Y. Sugiyama, Y. Tomita,
K. Ideno, I. Nishinaka, B. J. Qi, H. J. Kim, A. Iwamoto and T. Ohtsuki,
Phys. Rev. C \textbf{46}, 1922 (1992).
\bibitem{Caraley2000} A. L. Caraley, B. P. Henry, J. P. Lestone, and R.
Vandenbosch, Phys. Rev. C \textbf{62}, 054612-1 (2000).
\bibitem{Fabris1994} D. Fabris, E. Fioretto, G. Viesti, M. Cinausero, N. Gelli,
K. Hagel, F. Lucarelli, J. B. Natowitz, G. Nebbia, G. Prete, and R. Wada,
Phys. Rev. C \textbf{50}, R1261 (1994).
\bibitem{Morton1995} C. R. Morton, D. J. Hinde, J. R. Leigh, J. P. Lestone,
M. Dasgupta, J. C. Mein, J. O. Newton, and H. Timmers, Phys. Rev. C
\textbf{52}, 243 (1995).
\bibitem{Bohr1939} N. Bohr and J. A. Wheeler, Phys. Rev. \textbf{56}, 426
(1939).
\bibitem{Hagino1999} K. Hagino, N. Rowley and A. T. Kruppa, Comput. Phys. Commun.
\textbf{123}, 143 (1999).
\bibitem{Sierk1986} A. J. Sierk, Phys. Rev. C \textbf{33}, 2039 (1986).
\bibitem{Myers1966} W. D. Myers and W. J. Swiatecki, Nucl. Phys. \textbf{81}, 1
(1966).
\bibitem{Ignatyuk1975} A. V. Ignatyuk, M. G. Itkis, V. N. Okolovich, G. M.
Smirenkin, and A. Tishin, Yad. Fiz. \textbf{21}, 485 (1975) [Sov. J. Nucl. Phys.
\textbf{21}, 255 (1975).]
\bibitem{Reisdorf1981} W. Reisdorf, Z. Phys. A \textbf{300}, 227 (1981).
\bibitem{BBM1974} S. Bj\o rnholm, A. Bohr and B.R. Mottelson, Proc. Int. Conf.
on the Physics and Chemistry of Fission, Rochester 1973 (IAEA Vienna 1974) Vol.
1, p. 367.
\bibitem{Ignatyuk1985} A.V. Ignatyuk, G.N. Smirenkin, M.G. Itkis, S.I. Mul'gin
and V.N. Okolovich, Fiz. Elem. Chastits At. Yadra 16, 709 (1985) [Sov. J. Part.
Nucl. \textbf{16} (1985) 307].
\bibitem{Zagrebaev2001} V. I. Zagrebaev, Y. Aritomo, M. G. Itkis, Y. T.
Oganessian, and M. Ohta, Phys. Rev. C \textbf{65}, 014607 (2001).
\bibitem{Junghans1998} A. R. Junghans, M. de Jong, H.-G. Clerc, A. V. Ignatyuk,
G. A. Kudyaev, and K.-H. Schmidt, Nucl. Phys. A \textbf{629}, 635 (1998).
\bibitem{Ohta2001} M. Ohta, in Proceedings on Fusion Dynamics at the Extremes,
Dubna, 2000, edited by Yu. Ts. Oganessian and V. I. Zagrebaev, World Scientific,
Singapore, 2001, p. 110.
\bibitem{KBanerjee2017} K. Banerjee, Pratap Roy, Deepak Pandit, Jhilam Sadhukhan,
S. Bhattacharya, C. Bhattacharya, G. Mukherjee, T. K. Ghosh, S. Kundu, A. Sen,
T. K. Rana, S. Manna, R. Pandey, T. Roy, A. Dhal, Md. A. Asgar, and
S. Mukhopadhyay, Phys. Lett. B \textbf{772}, 105 (2017).
\bibitem{DPandit2018} Deepak Pandit, Srijit Bhattacharya, Debasish Mondal,
Pratap Roy, K. Banerjee, S. Mukhopadhyay, Surajit Pal, A. De, Balaram Dey,
and S. R. Banerjee, Phys. Rev. C \textbf{97}, 041301(R) (2018).
\bibitem{Lestone1999} J. P. Lestone, Phys. Rev. C \textbf{59}, 1540 (1999).
\bibitem{Kramers1940} H. A. Kramers, Phys. (Amsterdam, Neth.) \textbf{7}, 284
(1940).
\bibitem{JhilamPRC78} Jhilam Sadhukhan and Santanu Pal, Phys. Rev. C
\textbf{78}, 011603(R) (2008), Phys. Rev. C \textbf{79}, 01990(E) (2009).
\bibitem{Blocki1986} Blocki et al, Ann. Phys. \textbf{113}, 330 (1986).
\bibitem{Hofmann1997} H. Hofmann, Phys. Rep. \textbf{284}, 137 (1997).
\bibitem{Mukho1997} Tapan Mukhopadhyay and Santanu Pal, Phys. Rev. C \textbf{56}, 296 (1997).
\bibitem{Grange1983} P. Grang\'{e}, Li Jun-Qing, and H. A. Weidenm\"{u}ller,
Phys. Rev. C \textbf{27}, 2063 (1983). 
\bibitem{BhattPRC33} K.H. Bhatt, P. Grang\'{e}, and B. Hiller, Phys. Rev. C
\textbf{33}, 954 (1986).
\bibitem{HofmannPLB122}  H. Hofmann and J. R. Nix, Phys. Lett. \textbf{B122},
117 (1983).
\bibitem{Sarantites1978} D. G. Sarantites, L. Westerberg, R. A. Dayras, M. L.
Halbert, D. C. Hensley, and J. H. Barker, Phys. Rev. C \textbf{17}, 601 (1978).
\bibitem{Singh1992} N. L. Singh, S. Agarwal and J. Rama Rao, J. Phys. G: Nucl.
Part. Phys. \textbf{18}, 927 (1992).
\bibitem{Chakravarty1992} N. Chakravarty, P. K. Sarkar, and Sudip Ghosh,
Phys. Rev. C \textbf{45}, 1171 (1992).
\bibitem{Knoche1995} K. Knoche, L. L\"{u}demann, W. Scobel, B. Gebauer, D.
Hilscher, D. Polster, and H. Rossner, Phys. Rev. C \textbf{51}, 1908 (1995).
\bibitem{Singh2006} B. P. Singh, Manoj K. Sharma, M. M. Musthafa, H. D.
Bhardwaj, R. Prasad, Nucl. Inst. and Meth. A \textbf{562}, 717 (2006).
\bibitem{Pal1998} Santanu Pal and Tapan Mukhopadhyay, Phys. Rev. C \textbf{57},
210 (1998).
\bibitem{SarantitesPRC1978} D. G. Sarantites, L. Westerberg, M. L. Halbert,
R. A. Dayras, D. C. Hensley, and J. H. Barker, Phys. Rev. C \textbf{18}, 774
(1978).
\bibitem{Blann1974} M. Blann, Nucl. Phys. \textbf{A235}, 211 (1974).
\bibitem{Westerberg1978} L. Westerberg, D. G. Sarantites, D. C. Hensley, R. A.
Dayras, M. L. Halbert, and J. H. Barker, Phys. Rev. C \textbf{18}, 796 (1978).
\bibitem{Vigdor1982} S. E. Vigdor, H. J. Karwowski, W. W. Jacobs, S. Kailas, 
P. P. Singh, F. Soga, and T. G. Throwe, Phys. Rev. C \textbf{26}, 1035 (1982).
\bibitem{Cabrera2003} J. Cabrera, Th. Keutgen, Y. El Masri, Ch. Dufauquez, 
V. Roberfroid, I. Tilquin, J. Van Mol, R. Rgimbart, R. J. Charity, J. B.
Natowitz, K. Hagel, R. Wada, and D. J. Hinde, Phys. Rev. C \textbf{68}, 034613
(2003).
\bibitem{Bowman1962} Harry R. Bowman, Stanley G. Thompson, J. C. D. Milton, and
Wladyslaw J. Swiatecki, Phys. Rev. \textbf{126}, 2120 (1962).
\bibitem{Loveland1967} W. D. Loveland, A. W. Fairhall, and I. Halpern,
Phys. Rev. \textbf{163}, 1315 (1967).
\bibitem{Halpern1971} I. Halpern, Ann. Rev. Nucl. Part. Sci \textbf{21}, 245
(1971).
\bibitem{Franklyn1978} C. B. Franklyn, C. Hofmeyer and D. W. Mingay, Phys. Lett.
\textbf{78B}, 564 (1978).
\bibitem{Brosa1983} U. Brosa and S. Grossmann, Z. Phys. A \textbf{310}, 177
(1983).
\bibitem{Hwang1999} J. K. Hwang, A. V. Ramayya, J. H. Hamilton, W. Greiner,
J. D. Cole, G. M. Ter-Akopian, Yu. Ts. Oganessian, and A. V. Daniel (GANDS95
Collaboration), Phys. Rev. C \textbf{60}, 044616 (1999).
\bibitem{Kornilov2007} N. V. Kornilov, F.-J. Hambsch, A. S. Vorobyev, Nucl.
Phys. \textbf{A789}, 55 (2007).
\bibitem{Carjan2010} N. Carjan and M. Rizea, Phys. Rev. C \textbf{82}, 014617
(2010).
\bibitem{Capote2016} R. Capote, Chen Y.-J., F.-J. Hambsch, N. V. Kornilov,
J. P. Lestone, O. Litaize, B. Morillon, D. Neudecker, S. Oberstedt, T. Ohsawa,
Nucl. Data Sheets \textbf{131}, 1 (2016).
\bibitem{Eismont1965} V. P. Eismont, Sov. J. At. Energy \textbf{19}, 1000
(1965).
\bibitem{Skarsvag1973} K. Skarsvag, Phys. Scr. \textbf{7}, 160 (1973).
\bibitem{Hinde1984} D. J. Hinde, R. J. Charity, G. S. Foote, J. R. Leigh, J. O.
Newton, S. Ogaza, and A. Chatterjee, Phys. Rev. Lett. 52, 986 (1984).
\bibitem{Mahata2017} K. Mahata, and S. Kailas, Phys. Rev. C \textbf{95}, 054616
(2017).
\bibitem{Carjan2015} N. Carjan, M. Rizea, Phys. Lett. B \textbf{747}, 178
(2015).
\bibitem{Ye2010} W. Ye, Phys. Rev. C \textbf{81}, 054609 (2010).
\bibitem{Neeraj2017} Neeraj Kumar, Shabnam Mohsina, Jhilam Sadhukhan, and
Shashi Verma, Phys. Rev. C \textbf{96}, 034614 (2017).
\bibitem{Vardaci1998} E. Vardaci, G. La Rana, A. Brondi, R. Moro, A. Principe,
D. Fabris, G. Nebbia, G. Viesti, M. Cinausero, E. Fioretto et al., Eur. Phys. J.
A \textbf{2}, 55 (1998).
\bibitem{Lestone1993} J. P. Lestone, J. R. Leigh, J. O. Newton, D. J. Hinde,
J. X. Wei, J. X. Chen, S. Elfstrom, M. Zielinska-Pfabe, Nucl. Phys.
\textbf{A559}, 277 (1993).
\bibitem{Ikezoe1990} H. Ikezoe, N. Shikazono, Y. Nagame, Y. Sugiyama, Y. Tomita,
K. Ideno, A. Iwamoto, and T. Ohtsuki, Phys. Rev. C \textbf{42}, 342 (1990).
\bibitem{BeckermanPRL1979} M. Beckerman and M. Blann, Phys. Rev. Lett.
\textbf{42}, 156 (1979).
\bibitem{Blann1980} M. Blann, Phys. Rev. C \textbf{21}, 1770 (1980).
\bibitem{Lestone1991} J. P. Lestone, J. R. Leigh, J. O. Newton, D. J. Hinde,
J. X. Wei, J. X. Chen, S. Elfstrom, and D. G. Popescu, Phys. Rev. Lett.
\textbf{67}, 1078 (1991).
\bibitem{Wile1995} J. L. Wile, D. L. Coffing, E. T. Bauer, A. L. Michael, M. A.
Doerner, S. P. Baldwin, B. M. Szabo, B. Lott, B. M. Quednau, J. T\~{o}ke, W. U.
Schr\"{o}der and R. T. de Souza, Phys. Rev. C \textbf{51}, 1693 (1995).
\bibitem{Kalandarov2012} Sh. A. Kalandarov, G. G. Adamian and N. V. Antonenko, 
EPJ Web Conf. \textbf{38}, 09002 (2012).
\bibitem{Aleshin1993} V. P. Aleshin, J. Phys. G: Nucl. Part. Phys. \textbf{19},
307 (1993).
\end{thebibliography}
\end{document}